\documentclass[%
 aip,
 amsmath,amssymb,
 reprint,%
]{revtex4-1}

\usepackage{graphicx}
\usepackage{dcolumn}
\usepackage{bm}

\usepackage[utf8]{inputenc}
\usepackage[T1]{fontenc}
\usepackage{mathptmx}
\usepackage{etoolbox}
\usepackage{graphicx}
\usepackage{dcolumn}
\usepackage{bm}
\usepackage{color}
\usepackage[T1]{fontenc} 
\usepackage[english]{babel}
\usepackage{hyperref} 
\usepackage{booktabs} 
\usepackage{verbatim}
\usepackage{mathtools} 
\usepackage{caption}
\usepackage{subcaption}
\usepackage{pict2e}
\usepackage{lipsum} 
\usepackage{todonotes} 
\usepackage{siunitx}
\usepackage{physics}
\usepackage{adjustbox}
\usepackage{makecell}

\newcommand{\Rnum}[1]{\uppercase\expandafter{\romannumeral #1\relax}}

\draft 

\begin{document}


\title{Parametric study on ion acceleration from the interaction of ultra-high intensity laser pulses with near-critical density gas targets} 



\author{V. Ospina-Boh\'{o}rquez}\email{valeria.ospina@cea.fr}
\affiliation{CEA, DAM, DIF, F-91297 Arpajon, France}
\affiliation{Universit\'e Paris-Saclay, CEA, LMCE, 91690 Bruy\`eres-le-Ch\^atel, France}
\affiliation{Universit\'{e} de Bordeaux, CNRS, CEA, CELIA (Centre Lasers Intenses et Applications), UMR 5107, Talence, France}
\affiliation{Universidad de Salamanca, Salamanca, Spain}

\author{A. Debayle}\email{arnaud.debayle@cea.fr}
\affiliation{CEA, DAM, DIF, F-91297 Arpajon, France}
\affiliation{Universit\'e Paris-Saclay, CEA, LMCE, 91690 Bruy\`eres-le-Ch\^atel, France}

\author{J.~J. Santos}
\affiliation{Universit\'{e} de Bordeaux, CNRS, CEA, CELIA (Centre Lasers Intenses et Applications), UMR 5107, Talence, France}

\author{L. Volpe}
\affiliation{Universidad de Salamanca, Salamanca, Spain}
\affiliation{C.L.P.U. (Centro de L\'{a}seres Pulsados), Salamanca, Spain}

\author{L.~ Gremillet}\email{laurent.gremillet@cea.fr}
\affiliation{CEA, DAM, DIF, F-91297 Arpajon, France}
\affiliation{Universit\'e Paris-Saclay, CEA, LMCE, 91690 Bruy\`eres-le-Ch\^atel, France}

\begin{abstract}
 We present a parametric study based on 1-D particle-in-cell (PIC) simulations conducted with the objective of understanding the interaction of intense lasers with near-critical non-uniform density gas targets. Specifically, we aim to find an optimal set of experimental parameters regarding the interaction of a $\lambda_L = \SI{0.8}{\micro\metre}$, $I_L =10^{20}$ W/cm$^2$ ($a_0 = 8.8$), $\tau_L = \SI{30}{\femto\second}$ laser pulse with a near-critical non-uniform pure nitrogen gas profile produced by a non commercial gas nozzle. The PIC code Calder developed at CEA was used, and both the maximum electron density and the direct laser contribution to ion acceleration were studied. Shock formation was achieved for a peak electron density $n_e$ ranging between 0.35 $n_c$ and 0.7 $n_c$. In this density interval, the survival of a percentage of the laser pulse until the gas density peak, while being strongly absorbed ($>$90$\%$) and creating a hot electron population in the gas up-ramp, is singled out as a necessary condition for shock formation. Moreover, the laser absorption must give rise to a super ponderomotive heating of the target electrons in order to launch an electrostatic shock inside the plasma. The direct laser effect on ion acceleration consists in a strong initial density perturbation that enhances charge separation while the electron pressure gradients are identified as fundamental for shock formation. The production of a controlled and repetitive gas profile as well as the possibility of performing measurements with statistical meaning are highlighted as fundamental for conducting a thorough experimental study.

\end{abstract}


\maketitle 

\section{Introduction}

The generation of energetic ion particle sources, with energies in the 10-$\SI{100}{\mega\electronvolt}$ range, is one of the major applications of ultra-intense short-pulse lasers, made available thanks to the development of the chirped pulse amplification (CPA) technique.\cite{Strickland_1985} These particle sources are employed in ultra-fast probing methods in science and industry, such as proton radiography\cite{Borguesi_2003} and time-resolved probing of laser-created transient states.\cite{Borguesi_2004,Santos_2015,Ehret_2017} These particles are also used in the creation of warm dense matter,\cite{Roth_2009} as well as in isochoric heating of dense plasmas\cite{Patel_2003} and radioisotope and intense neutron sources production with possible medical spin offs.\cite{Fritzler_2003,Spencer_2001} Most of these applications exploit the unique properties of laser-driven ion beams, notably their short duration, high number density, low emittance, high laminarity and compactness compared to conventional particle accelerators as well as the possibility of a high repetition rate (HRR) production when working at modern fs-laser facilities.

The search for more energetic and efficient laser-driven particle sources motivates the investigation of novel acceleration schemes involving new states of matter, target shapes and compositions. The laser-matter interaction is divided in two regimes depending on the transparency/opacity properties of the irradiated (and ionized) medium. The plasma density threshold is governed by the so-called critical density $n_c = 1.11\times10^{21}/\lambda_L^2 \ [\mathrm{cm^{-3}}]$ where $\lambda_L$ is the laser wavelength in $\SI{}{\micro\metre}$. A medium providing an electron density $n_e$ (upon ionization) will be opaque if  $n_e > n_c = 1.74 \times 10^{21} \ \mathrm{cm^{-3}}$ considering a $\lambda_L = \SI{0.8}{\micro\metre}$ laser. Opaque plasmas are know as overdense plasmas ($n >> n_c$) while transparent ones are given the name of underdense plasmas ($n << n_c$). Solid and liquid targets are normally opaque to the laser. However, the laser interaction with the outer layer of the material allows for a $\approx1 - 50\%$ energy transfer to the target electrons accelerating them to $\approx$MeV energies. These fast electrons induce strong ($\approx\mathrm{TV/m}$) charge separation fields over $\SI{}{\micro\metre}$ scales once propagated through the target and crossing the target-vacuum interface. Ions with the higher charge-to-mass ratios are preferentially accelerated by the rear-side sheath field that has been created, referring normally to protons present as organic impurities in the material's surface. This process is called Target Normal Sheath Acceleration (TNSA),\cite{Macchi_2013,Daido_2012} a term coined in the early 2000s to name the general and long-known process of sheath-field ion acceleration arising in non-uniform plasmas. Broad energy spectrum proton beams of a maximum energy of $\SI{85}{\mega\electronvolt}$\cite{Wagner_2016} have been produced following this scheme. 

When the medium's density drops strongly below $n_c$, the laser is allowed to propagate inside the material creating non-linear wakefields capable of accelerating electrons to relativistic velocities. At the same time, the laser ponderomotive force expels electrons creating a channel along the propagation axis, which expands radially. Coulomb explosion of the local ions is then possible after the ponderomotive expulsion of the electrons which leaves the ions to strongly experience the unscreened field arising from charge separation. TNSA type ion acceleration can also arise from the expelled electrons of the expanding sheath front. Moreover, the return electron current that is formed once exiting into vacuum sustains a strong quasistatic magnetic field in the vacuum interface which can also efficiently accelerate as well as collimate ions.\cite{Bulanov_2007,Bulanov_2005,Park_2019,Nakamura_2010} Multi-MeV ion beams have been produced in this interaction regime. \cite{Krushelnick_1999,Kuznetsov_2001,Willingale_2006}

Fewer studies have addressed the case of near-critical plasmas, i.e. of electron density approaching the critical density, $n_e \approx n_c$, due to the technical difficulty of achieving such densities in a controlled and repetitive manner. When $n_e \approx n_c$, intense laser pulses can propagate over large distances and volumetrically heat the plasma electrons to relativistic temperatures. The interaction of two collisionless plasma regions with different densities, temperatures or velocities can then lead to the formation of non-linear electrostatic shocks capable of trapping and/or reflecting charged particles. These shocks can be triggered in several ways: by applying a laser piston to the overcritical boundary of a uniform overcritical plasma;\cite{Denavit_1992,Silva_2004} or in partially transparent plasmas where strong electron pressure gradients, shorter than the local Debye length $\lambda_D$, and/or laser-driven ion acceleration can trigger shock formation. Near-critical plasmas are predicted to give rise to new ion acceleration regimes combining TNSA and collisionless shock acceleration (CSA),\cite{Fiuza_2012,Fiuza_2013,Dieckmann_2013a,Dieckmann_2013b,Lecz_2015,Liu_2016,Liu_2016,Liu_2018} as well as enhanced hot-electron production, that is, beyond the standard ponderomotive scaling, i.e. the mean hot electron energy $\epsilon_h \approx \sqrt{1+{a_0}^2/2}-1$, where $a_0=e E_0/m_e c\omega_{0}$ is the normalized laser amplitude and $E_0$ the laser electric field in vacuum, $m_e$ is the electron rest mass and $c$ the speed of light in vacuum. Our Calder code\cite{Lefebvre_2003} preliminary simulations predict efficient volumetric electron heating due to phase mixing between bulk electrons and electrons trapped in laser-induced plasma waves.\cite{Debayle_2017} The strong electron pressure gradients that result in inhomogenous plasmas are able to trigger an electrostatic shock,\cite{Moiseev_1963,Forslund_1970,Sorasio_2006,Cairns_2014} favoring and enhancing ion acceleration.\cite{Debayle_2017} Furthermore, CSA is expected to produce more peaked ion spectra in comparison to the broad spectra produced through TNSA.\cite{Boella_2018} Previous experimental attempts in this direction have resorted to the use of double-layer targets composed of a near-critical layer (e.g. carbon nanotubes) and a thin diamond-like solid layer \cite{Bin_2015,Bin_2018,Ma_2019,Pazzaglia_2020} with $\lambda_L = \SI{0.8}{\micro\metre}$ fs lasers, or exploded foils used in double pulse ns/ps configuration with $\lambda_L = \SI{1}{\micro\metre}$ lasers.\cite{Antici_2017,Pak_2018,Rosmej_2019} Moderate-density gas jets irradiated by $\mathrm{CO_2}$ ($\SI{10}{\micro\metre}$ wavelength) laser pulses,\cite{Haberberger_2012,Tresca_2015,Tsung_2012} where the critical density constraint is eased, have also been employed and the laser pedestal has been used to sculpt the gas up-ramp into a steep density front.\cite{Chen_2017,Puyuelo-Valdes_2019} The advent of new high-intensity HRR facilities like Apollon (France), VEGA-3 (Spain), L4 Aton (Czech Republic) or BELLA PW (USA) drives the application of this ion acceleration scheme using high-density gas-jets, a HRR-compatible debris-free targetry. Pioneer experiments have been conducted both in the TW \cite{Sylla_2013,Ehret_2020} and the PW\cite{Singh_2020} regime with preliminar although encouraging ion acceleration results.

In this paper, we will examine collisionless electrostatic shock (CES) formation in transparent plasmas under conditions accessible to a laser system characterized by $I_L$=1.7 x $10^{20}$ W/cm$^2$, $\lambda_L=\SI{0.8}{\micro\metre}$ and $\tau_L = \SI{30}{\femto\second}$ (parameters of the VEGA-3 laser, CLPU, Spain), and considering as a target a highly non-uniform near-critical-density gas jet produced by the a non-commercial supersonic nozzle \cite{Puyuelo-Valdes_2020}, plotted in Fig. \ref{fig:Density_lineouts}. In the case of a $\mathrm{N_2}$ gas, CES are found to occur within specifically, the quite a narrow electron density $n_e$ range between 0.35 $n_c$ and 0.7 $n_c$. The transmission of a significant fraction of the laser pulse across the density peak, while being strongly absorbed (80$\%$ - 90$\%$) and creating a hot electron population in the gas up-ramp, is identified as a necessary condition for CES formation. The latter stems from the strong electron pressure gradients located in the density down-ramp. The direct laser effect on ion acceleration is a strong initial density perturbation that enhances charge separation while electron pressure gradients are identified as the main cause of shock formation. The addition of lighter helium ions to the $\mathrm{N_2}$ gas leads to peak in the velocity spectra, which may constitute a signature of CES formation. Finally, the production of a controlled and repetitive gas profile as well as the possibility of performing measurements with statistical meaning are highlighted as fundamental for conducting a thorough experimental study.

This paper is organized as follows. The simulation parameters and initial conditions are described in Section \ref{Sim_param}. The electron energization and plasma heating processes are presented in Section \ref{Electron_accel}. The description of the different ion acceleration mechanisms including CSA and TNSA is exposed in Section \ref{Ion_accel}. Finally, the conclusions are summarized in Section \ref{Conclusions}.

\section{Simulation parameters and initial conditions}
\label{Sim_param}

The present 1-D simulations describe the interaction of a laser pulse characterized by a normalized field strength $a_0$=8.8 ($I_L$=1.7 x $10^{20}$ W/cm$^2$), a laser wavelength $\lambda_L=\SI{0.8}{\micro\metre}$ and a $\tau_L = \SI{30}{\femto\second}$ FWHM duration, corresponding to the VEGA-3 laser parameters. The laser is linearly polarized along the $y$-axis and it is injected along the $x$-axis from the left-side of the simulation box. This study is restricted to a 1-D geometry because of the large spatial ($\approx$mm) and temporal ($\approx\SI{10}{\pico\second}$) scales considered, and the relatively fine resolution dictated by the ~$n_c$ maximum density of the fully ionized $\mathrm{N_2}$ gas profile. Field and impact ionization together with elastic Coulomb collisions between all charged particles were considered. Absorbing boundary conditions were applied for fields and particles. Each cell initially contained 100 ions of each species. The initial ion temperature was $T_i = \SI{1}{\electronvolt}$. The temporal and spatial resolutions were $\Delta t$= 0.095 $\omega_0^{-1}$ ($\SI{0.05}{\femto\second}$) and $\Delta x$= 0.1 $c/\omega_0$ ($\SI{0.016}{\micro\metre}$), respectively. The modeled gaseous targets were composed of either atomic nitrogen or a mixture of 90$\%$ $\mathrm{N_2}$ and 10$\%$ He.

The initial gas density profile was obtained by performing fluid dynamics simulations with the FLUENT code \cite{Fluent_2015}. The reference profile was that predicted with the S900 shock nozzle \cite{Puyuelo-Valdes_2020} and a backing $\mathrm{N_2}$ gas pressure of $\SI{400}{\bar}$, measured far enough from the nozzle to avoid damaging it.\cite{Ehret_2020} This profile, plotted in Fig. \ref{fig:Density_lineouts} (blue curve), is characterized by a very narrow ($\approx\SI{15}{\micro\metre}$) density peak (up to $n_{e,max}=9.8 \ n_c$) sitting on broad ($\approx\SI{1}{\milli\metre}$), quasi-exponential symmetric wings. The chosen gas is nitrogen in order to replicate in a simpler manner the experimental conditions.

In the following and unless explicitly marked differently the density, velocity, time, distance, mass, electron momentum, ion momentum, energy and electric field are normalized to the following quantities, respectively:

\begin{equation}
n_c, c, \omega_{0}^{-1}, {c}/{\omega_{0}}, m_e, m_ec, m_ic, m_ec^2, m_e\omega_{0}c/e,
\label{eq:norm}   
\end{equation}

\noindent
where $\lambda_{0}=2\pi/k_0=2\pi c/\omega_0$ is taken to be equal to \SI{1}{\micro\metre}, $n_c$ is the critical density for a 1 $\mu$m laser wavelength equal to 1.11$\times 10^{27} \ \mathrm{m^{-3}}$, $c$ is the velocity of light in vacuum,  $\omega_{0}^{-1}$ is the corresponding inverse laser frequency equal to $\SI{0.53}{\femto\second}$, $e$ is the electron charge and $m_e$ and $m_i$ are the electron and ion rest masses, respectively.

Starting from the reference gas profile we investigate the laser-driven plasma dynamics in uniformly rescaled $\mathrm{N_2}$ profiles with $n_{e,max}$ = 2.1, 0.7, 0.35 and 0.14 $n_c$. Several runs were  performed to examine the role of the sole hot electrons to drive the CES. Additional runs were  performed to examine the changes brought to the interaction by a gas mixture of $90 \% \ \mathrm{N_2}$ and $10\% \ $He or $90 \% \ \mathrm{N_2}$ and $10\% \ $H, as well as the effect of interacting with a stretched laser pulse.

Table \ref{tab:summary_simulations} summarizes the simulations' main parameters regarding both the laser and gas target. Note that following the normalization given in Eq. \ref{eq:norm}, $a_0 = 8.8$ for a $\lambda_L = \SI{0.8}{\micro\metre}$ transforms into $a_{0,norm.} = 11$ given the normalization for  $\lambda_L = \SI{1}{\micro\metre}$, as written in Table \ref{tab:summary_simulations}.

\begin{table}[ht!]
		\begin{tabular}{ | c | c | c | c | c | c |c | }
			\hline
			\thead{Sim. \\ number} & \thead{$n_{at,max}$ \\ $[n_c]$} & \thead{$n_{e,max}$ \\ $[n_c]$} & Gas type & \thead{Laser\\ ON/OFF} & $a_0$ & \thead{$\tau_L$ \\ $[\omega_0^{-1}]$}\\
			\hline
			\Rnum{1} & 1.4 & 9.8 & $\mathrm{N_2}$ & ON & 11 & 56\\
			\Rnum{2} & 0.3 & 2.1 & $\mathrm{N_2}$ & ON & 11 & 56\\
			\Rnum{3} & 0.1 & 0.7 & $\mathrm{N_2}$ & ON & 11 & 56\\
			\Rnum{4} & 0.05 & 0.35 & $\mathrm{N_2}$ & ON & 11 & 56\\
			\Rnum{5} & 0.02 & 0.14 & $\mathrm{N_2}$ & ON & 11 & 56\\
			\Rnum{6} & 0.1 & 0.7 & $\mathrm{N_2}$ & OFF ($T_e = a_0$) & $-$ & $-$\\
			\Rnum{7} & 0.1 & 0.7 & $\mathrm{N_2}$ & ON & 6.3 & 169\\
			\Rnum{8} & 0.1 & 0.7 & $90 \% \ \mathrm{N_2}$+$10\% \ $He & ON & 11 & 56\\	
			\Rnum{9} & 0.1 & 0.7 & $90 \% \ \mathrm{N_2}$+$10 \% \ \mathrm{H_2}$ & ON & 11 & 56\\
			\Rnum{10} & 0.1 & 0.7 & $\mathrm{N_2}$ & OFF ($T_e = a_0/2$) & $-$ & $-$\\	
			\hline	
	\end{tabular}
	\caption{\label{tab:summary_simulations} Main parameters of the 1-D simulations.}
\end{table}

\section{Electron acceleration and plasma heating}
\label{Electron_accel}

\subsection{Areal density needed for complete laser absorption}

To discern between high and moderate/low density simulations the criterion of complete laser absorption inside the gas profile is considered. It is important to differentiate between these two interaction regimes since they will trigger different electron acceleration mechanisms, as will be discussed later on. This condition has previously been studied in Ref. \cite{Debayle_2017}, where the minimum areal density required for complete laser absorption was found to scale approximately as

\begin{equation}
\sigma_{abs} = \int_{x_0}^{x_{abs}}dx n_e(x)\approx\frac{a_0^2\tau_L}{2\langle
	\gamma_e\rangle} 
\label{eq:sigma_abs}
\end{equation}

\noindent
where $\langle
\gamma_e\rangle \approx a_0$ is the relativistic Lorentz factor corresponding to the mean bulk electron energy.

\begin{figure}[ht]\centering
	\includegraphics[width=\columnwidth]{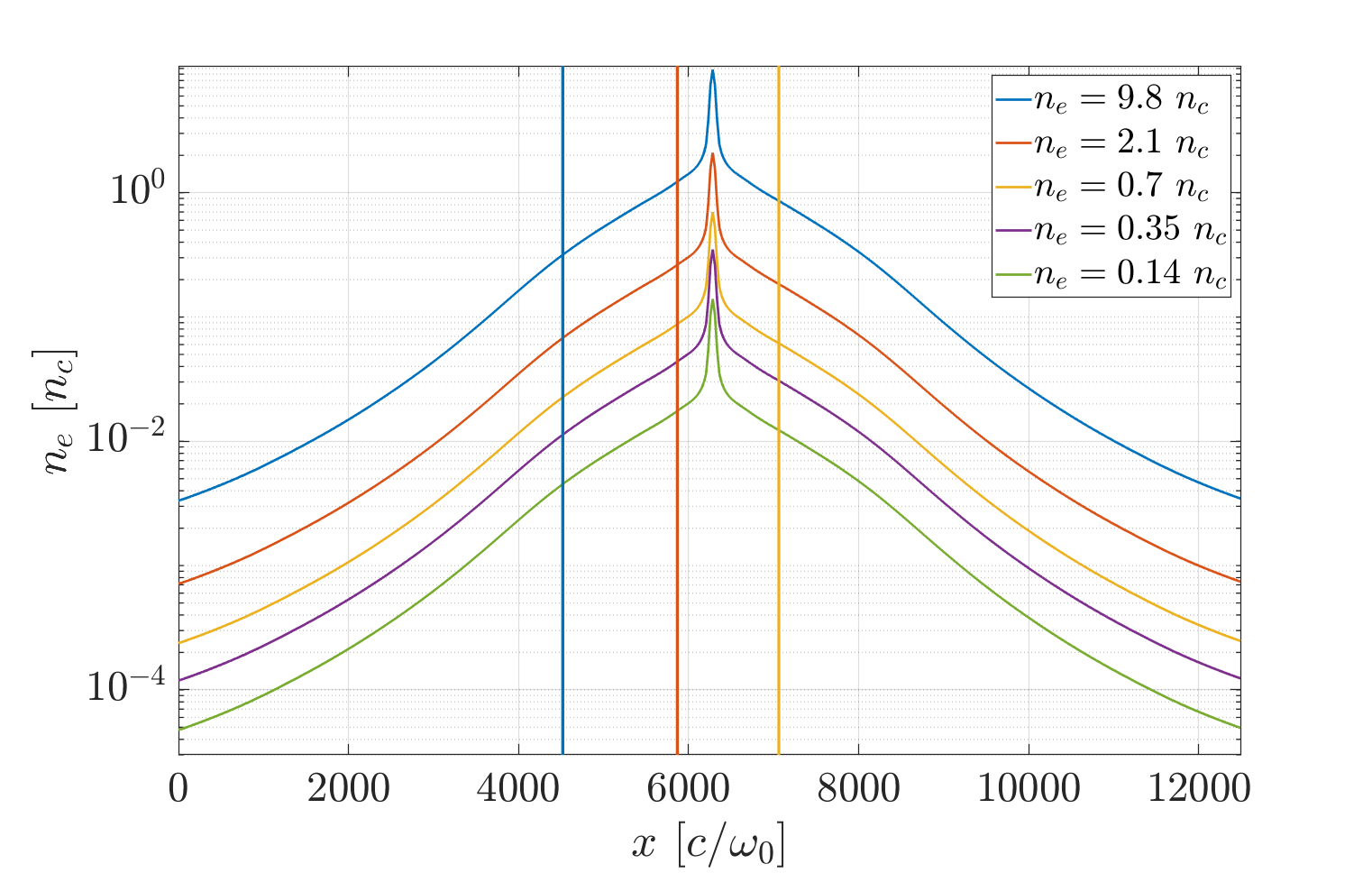}
	\caption{Electron density $n_e$ profiles input in the 1-D PIC simulations.}
	\label{fig:Density_lineouts}
\end{figure}

\begin{figure*}[ht!]
	\centering
	\begin{subfigure}[b]{0.49\linewidth}
		\centering
		\includegraphics[width=\linewidth]{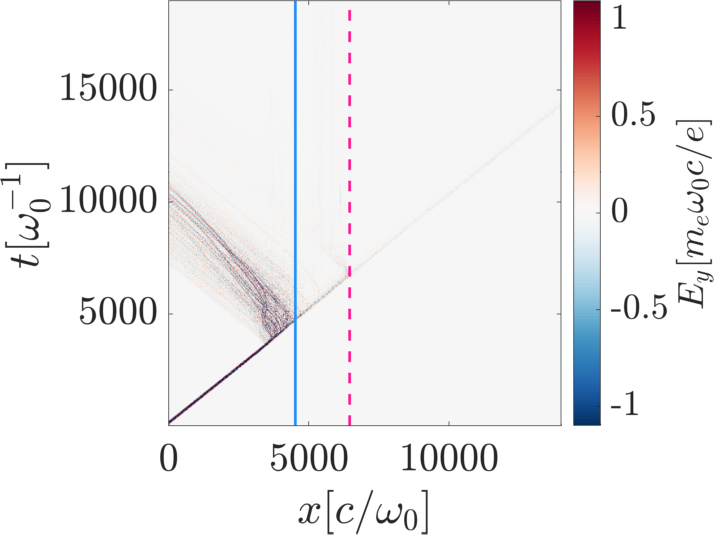}
		\begin{picture}(20,20)
		\put(-40,215){\large Simulation \Rnum{1}: $n_{e} = 9.8 \ n_c$}
		\put(-50,190){\large\textbf{(a)}}
		\end{picture}	
	\end{subfigure}
	\begin{subfigure}[b]{0.49\linewidth}
		\centering
		\includegraphics[width=\textwidth]{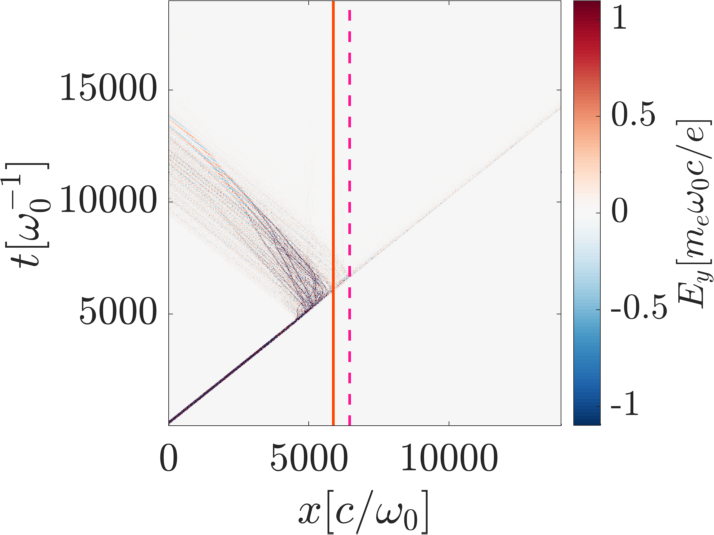}
		\begin{picture}(20,20)
		\put(-40,215){\large Simulation \Rnum{2}: $n_{e} = 2.1 \ n_c$}
		\put(-50,190){\large\textbf{(b)}}
		\end{picture}
	\end{subfigure}
	\begin{subfigure}[b]{0.49\linewidth}
		\centering
		\includegraphics[width=\textwidth]{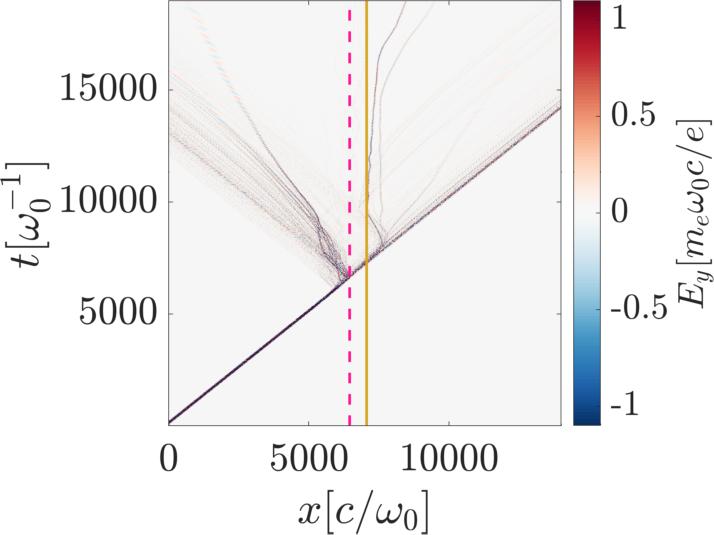}
		\begin{picture}(20,20)
		\put(-40,215){\large Simulation \Rnum{3}: $n_{e} = 0.7 \ n_c$}
		\put(-50,190){\large\textbf{(c)}}
		\end{picture}
	\end{subfigure}
	\begin{subfigure}[b]{0.49\linewidth}
		\centering
		\includegraphics[width=\textwidth]{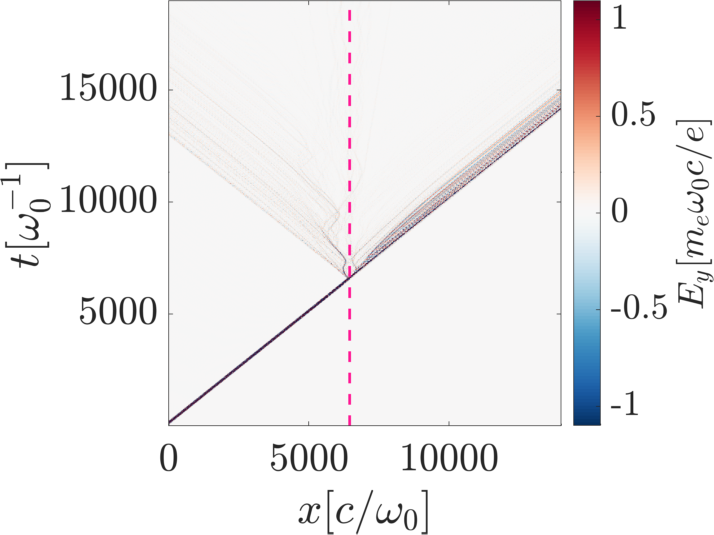}
		\begin{picture}(20,20)
		\put(-40,215){\large Simulation \Rnum{4}: $n_{e} = 0.35 \ n_c$}
		\put(-50,190){\large\textbf{(d)}}
		\end{picture}
	\end{subfigure}
	\caption{$(x,t)$ evolution of the laser field $E_y$ propagation in the (a) S.\Rnum{1}, (b) S.\Rnum{2}, (c) S.\Rnum{3} and (d) S.\Rnum{4}, see Table \ref{tab:summary_simulations}. The theoretical positions for complete laser absorption deduced from Eq. \ref{eq:sigma_abs}, are marked with solid lines and summarized in Table \ref{tab:table_laser_absorption}. The profile's peak density is marked by the pink dashed line at 6432 $c/\omega_0$.}
	\label{fig:Ey_charts_areal_density}
\end{figure*}
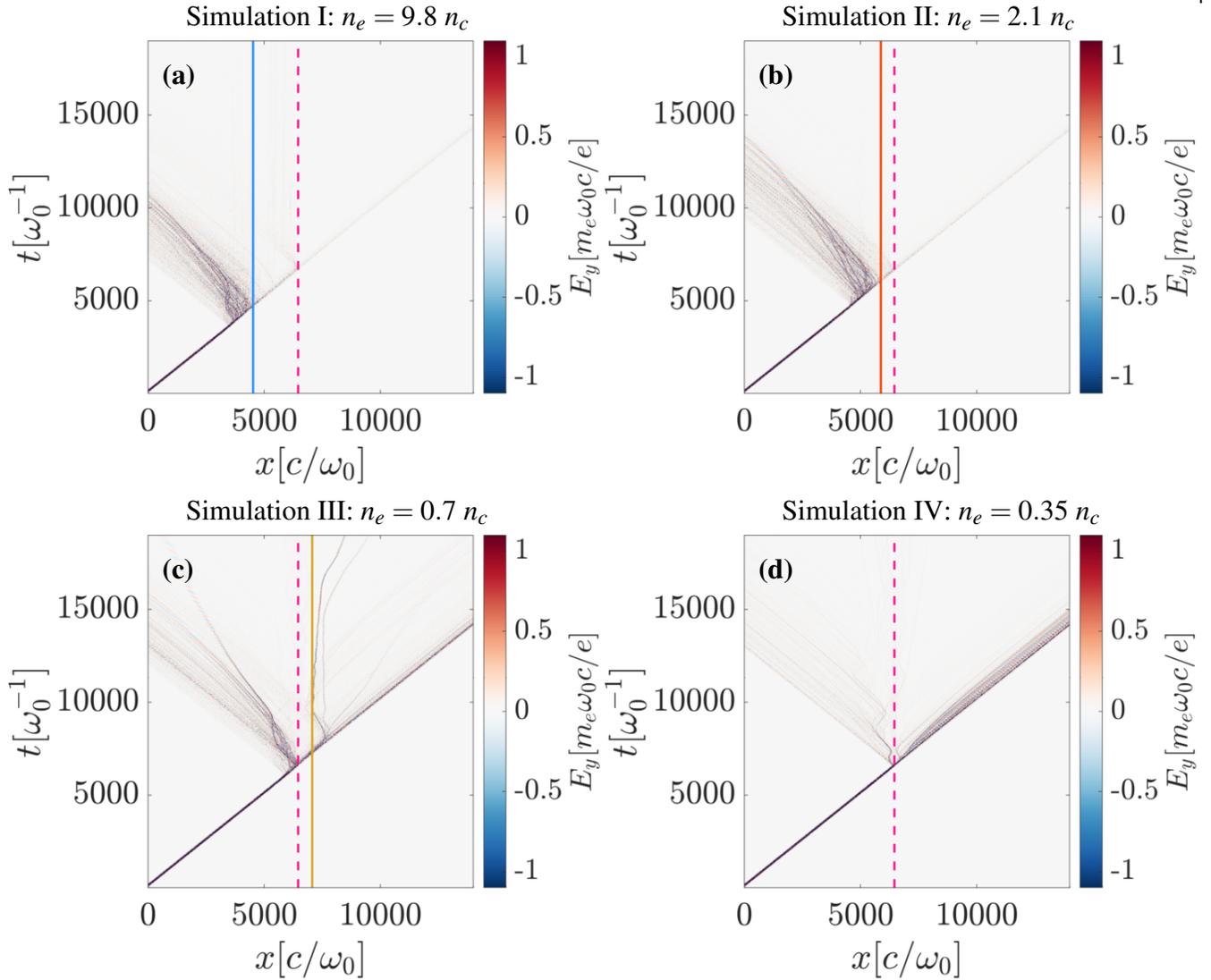

According to this criterion, a $I_L$=1.7 x $10^{20}$ W/cm$^2$, $\lambda_L=\SI{0.8}{\micro\metre}$, $\tau_L = \SI{30}{\femto\second}$ laser pulse should be fully absorbed before reaching the density peak in the simulations with $n_{e,max}\geq2.1 \ n_c$ (Sims. \Rnum{1}-\Rnum{3}). In the simulations with $n_{e,max}<2.1 \ n_c$ the laser is able to cross the entire profile. The crossed thicknesses of complete laser absorption $x_{abs}$ in Sims. \Rnum{1}-\Rnum{3} are indicated as solid lines in Figs. \ref{fig:Density_lineouts} and \ref{fig:Ey_charts_areal_density}, and the obtained $x_{abs}$ values are summarized in Table \ref{tab:table_laser_absorption}.

\begin{table}
	\centering
	\resizebox{0.15\textwidth}{!}{\begin{tabular}{| c | c | c |}
			\hline
			\thead{Sim. \\ number} & \thead{$n_{e,max}$\\ $[n_c]$} & \thead{$x_{abs}$\\ $[c/\omega_0]$}\\
			\hline
			\Rnum{1} & 9.8 & 4520\\
			\Rnum{2} & 2.1 & 5868\\
			\Rnum{3} & 0.7 & 7061\\
			\Rnum{4} & 0.35 & -\\
			\Rnum{5} & 0.14 & -\\
			\hline
	\end{tabular}}
	\caption{\label{tab:table_laser_absorption} $X$-coordinates where complete laser absorption is met for Sims. \Rnum{1}-\Rnum{3} considering full ionization of the nitrogen gas in Eq. \ref{eq:sigma_abs}. In Sims. \Rnum{4} and \Rnum{5} the laser is able to cross the entire profile without being completely absorbed.}
\end{table}

The $(x,t)$ evolution of the $y$-componen of the laser field $E_y$ is plotted for various maximum electron densities (9.8 $n_c$, 2.1 $n_c$, 0.7 $n_c$ and 0.35 $n_c$) in Figs. \ref{fig:Ey_charts_areal_density}a-d. The colormaps are saturated to highlight the transmitted fraction of the pulse. The density peak is located at $x = $ 6432 $c/ \omega_0$ and it is marked with pink dashed lines. The positions corresponding to full laser absorption, as predicted by Eq. \ref{eq:sigma_abs}, are plotted as solid lines. In the highest density case (Fig. \ref{fig:Ey_charts_areal_density}a) most of the laser has been absorbed or reflected at about 5000 $c/\omega_0$, approximately 1000 $c/\omega_0$ ahead of the density peak, a location consistent with Eq. \ref{eq:sigma_abs}. Only $0.1\%$ of the laser energy has been transmitted and a few percent of the laser intensity ($I$ $\propto  a_0^2$). 

In the high density cases, above $n_{at}\geq 2.1 \ n_c$ (Figs. \ref{fig:Ey_charts_areal_density}a and b), strong laser reflection takes place possibly as a consequence of backward stimulated Raman scatering (BSRS) \cite{Guerin_1995,Guerin_1996,Moreau_2017,Moreau_2018}. 

Between $0.35 \geq n_{e} \geq 0.7 \ n_c$, in the moderate density regime (Figs. \ref{fig:Ey_charts_areal_density}c and d), the pulse transmitted across the density peak remains intense enough ($a_0 \approx 3$ at $\omega_0t = \ 6500$ in Fig. \ref{fig:Ey_charts_areal_density}c) to induce a plasma wakefield and thus continue energizing the down-ramp bulk electrons. Several electromagnetic solitons are created as well but are not believed to have a strong effect on the electron acceleration, let alone the ion dynamics. 

\subsection{Stochastic electron heating}

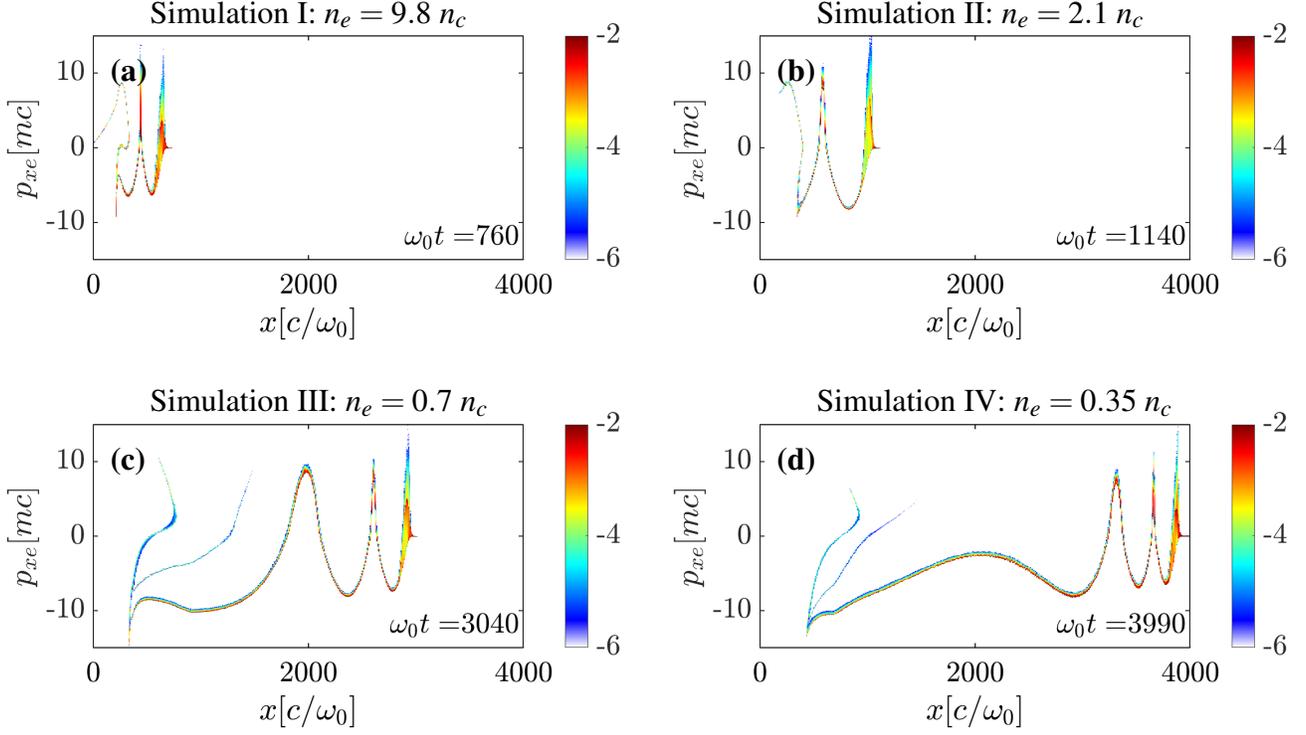
\begin{figure*}[ht!]
	\centering
	\begin{subfigure}[b]{0.49\linewidth}
		\centering
		\includegraphics[width=\linewidth]{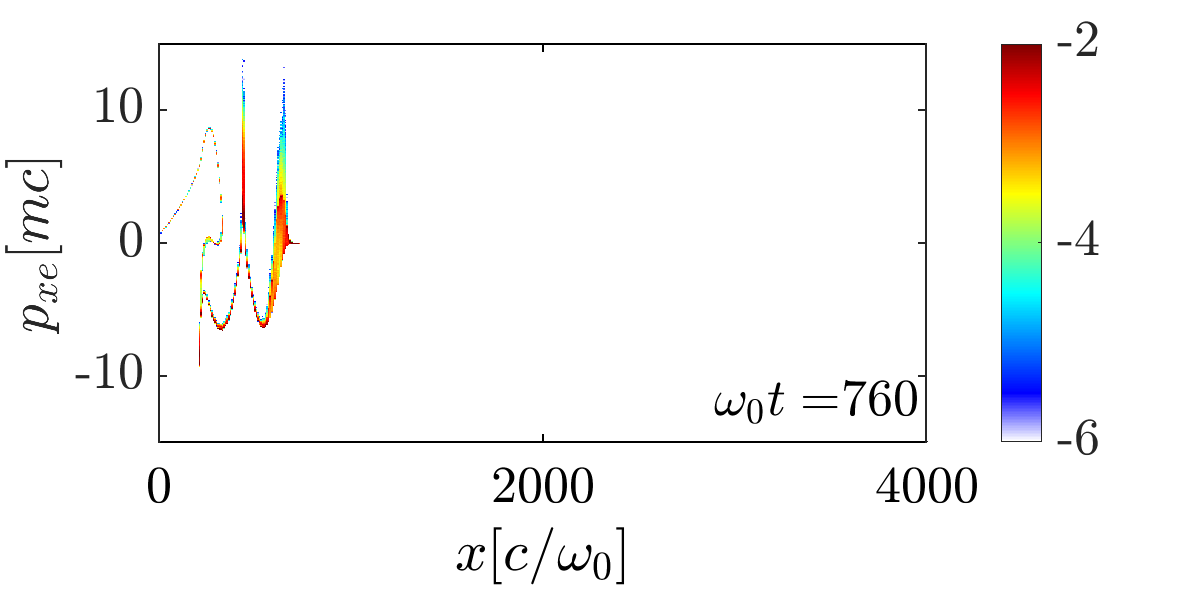}
		\begin{picture}(20,20)
		\put(-60,142){\large Simulation \Rnum{1}: $n_{e} = 9.8 \ n_c$}
		\put(-75,120){\large\textbf{(a)}}
		\end{picture}	
	\end{subfigure}
	\begin{subfigure}[b]{0.49\linewidth}
		\centering
		\includegraphics[width=\textwidth]{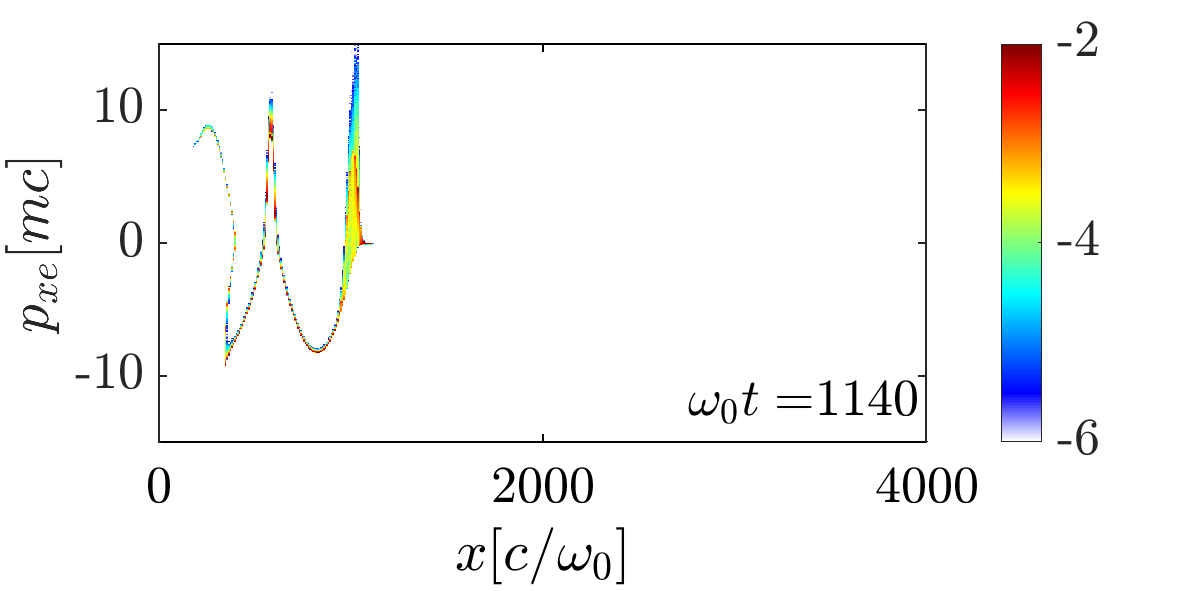}
		\begin{picture}(20,20)
		\put(-60,142){\large Simulation \Rnum{2}: $n_{e} = 2.1\ n_c$}
		\put(-75,120){\large\textbf{(b)}}
		\end{picture}
	\end{subfigure}
	\begin{subfigure}[b]{0.49\linewidth}
		\centering
		\includegraphics[width=\linewidth]{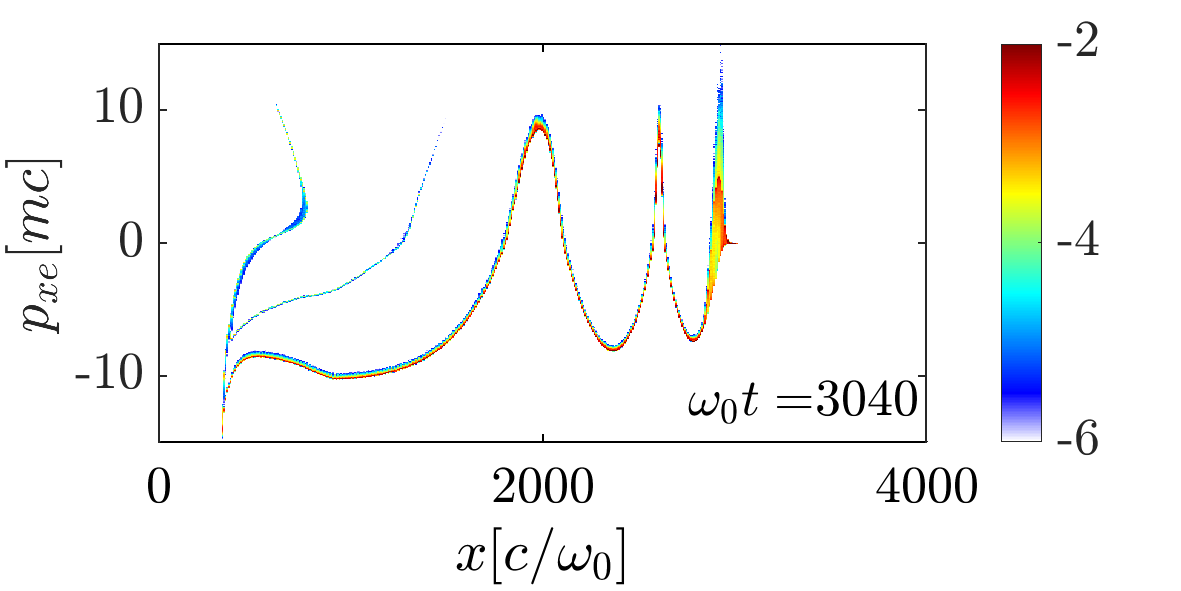}
		\begin{picture}(20,20)
		\put(-60,142){\large Simulation \Rnum{3}: $n_{e} = 0.7 \ n_c$}
		\put(-75,120){\large\textbf{(c)}}
		\end{picture}	
	\end{subfigure}
	\begin{subfigure}[b]{0.49\linewidth}
		\centering
		\includegraphics[width=\textwidth]{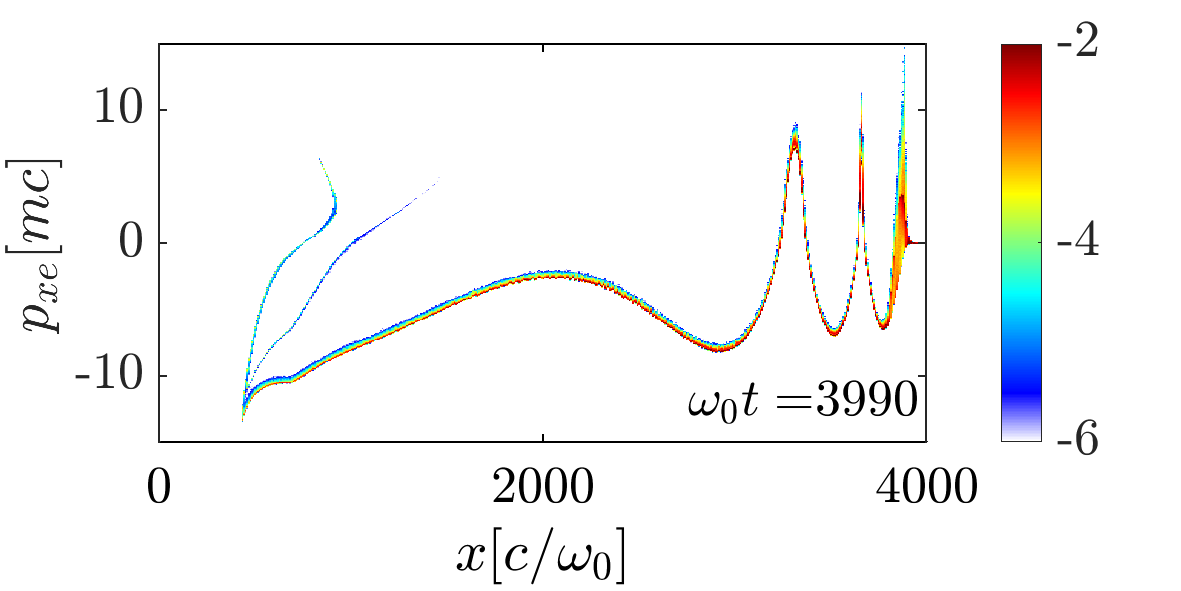}
		\begin{picture}(20,20)
		\put(-60,142){\large Simulation \Rnum{4}: $n_{e} = 0.35\ n_c$}
		\put(-75,120){\large\textbf{(d)}}
		\end{picture}
	\end{subfigure}
	\caption{{\color{black}Electron $(x, p_x)$ phase space for (a) S.\Rnum{1}, (b) S. \Rnum{2}, (c) S. \Rnum{3} and (d) S. \Rnum{4}, see Table \ref{tab:summary_simulations}, illustrating wakefield-type electron acceleration taking place in the low density part of the gas target up-ramp.}}
	\label{fig:Wakefield}
\end{figure*}

All simulations show that, early in the interaction, wakefield-type electron acceleration takes place in the fast-ionized plasma. This fluid stage of electron acceleration is shown in the electron  $(x, p_x)$ phase spaces of Figs. \ref{fig:Wakefield}a-d corresponding to Sims. \Rnum{1} - \Rnum{4}, see Table \ref{tab:summary_simulations}. The wakefield matching condition is met for $\tau_Lc \approx 4\pi c/\omega_{pe}$.\cite{Teychenne_1993} Thus, reducing the gas peak density will shift the resonance condition to a deeper region of the plasma. This is illustrated by the similarity of the electron $(x,p_x)$ phase spaces in Figs. \ref{fig:Wakefield}a-d, for raising positions in the plasma while reducing the plasma density.

\begin{figure*}[ht]\centering
	\includegraphics[width=0.7\textwidth]{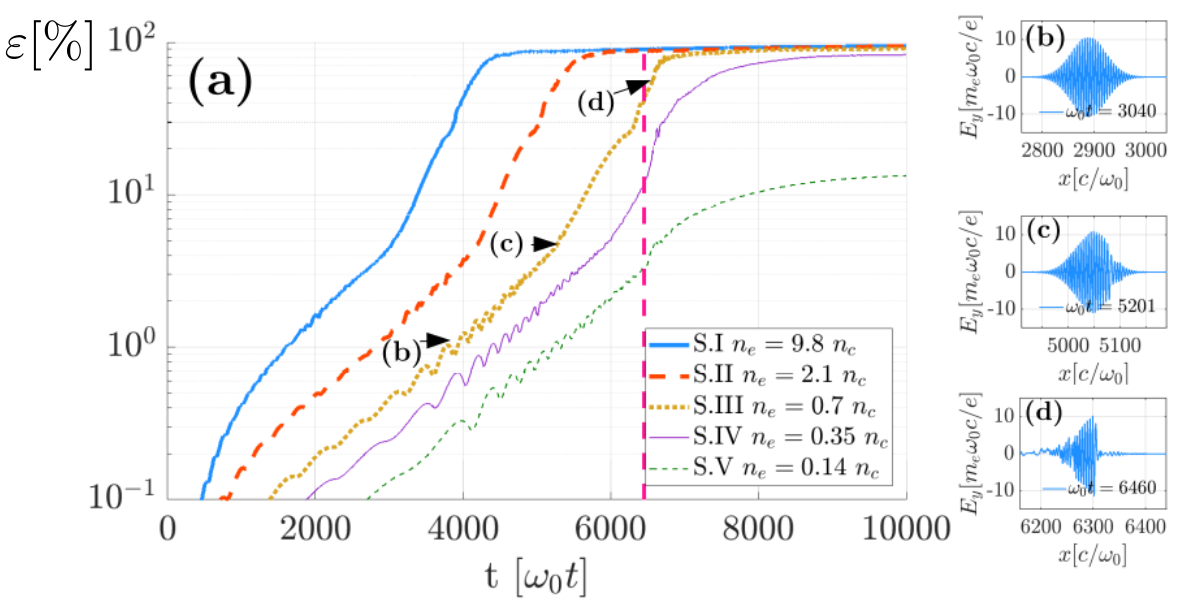}
	\caption{(a) Percentage of absorbed laser energy with respect to time for S.\Rnum{1} (blue solid line), S.\Rnum{2} (orange dashed line), S.\Rnum{3} (yellow squared-dotted line), S.\Rnum{4} (dark purple thin solid line) and S.\Rnum{5} (green thin dashed line), see Table \ref{tab:summary_simulations}. The instant when the laser pulse arrives to the density peak $\omega_0t = 6450$ is marked with a vertical pink dashed line. The insets of the figure correspond to the $E_y$ laser field in S.\Rnum{3} at times (b) $\omega_0t = 3990$, (c) $\omega_0t = 5201$ and (d) $\omega_0t = 6460$, marked with black arrows in (a).}
	\label{fig:ART_all}
\end{figure*}

At the beginning of the interaction, the wakefield electron energization stage plotted in Figs. \ref{fig:Wakefield}a-d entails a low laser energy absorption rate. The latter translates into the gentle first section of the laser absorption curves, plotted against time in Fig. \ref{fig:ART_all}a. Simulations \Rnum{1}-\Rnum{4} exhibit a transition from a laminar wakefield stage to a turbulent regime where the laser absorption rate strongly increases. The first stage of electron energization occurs until the times $\sim3000, 4200, 5000$ and $6000 \ \omega_0^{-1}$ for Sims. \Rnum{1}-\Rnum{4}, respectively. The steepening of the laser absorption curves coincides with the instant where the laser reaches the electron density $n_e$ that satisfies the wakefield matching condition. 

At the same time, since the laser pulse is longer than the local electron plasma wavelength $\lambda_{pe}$ ($\tau_Lc = 56 \ c/\omega_0 > \lambda_{pe} = 3 \ c/\omega_0$, in S.\Rnum{1}), the wakefield is generated in the front part of the laser pulse. The front of the laser pulse is then strongly depleted in the process of transferring its energy to the wakefield.\cite{Vieira_2010} This is observed in Figs. \ref{fig:ART_all}b-d, which correspond to the laser $E_y$ field of S.\Rnum{3} plotted at times $\omega_0t = $ 3990, 5201 and 6270. This so-called \textit{optical shock} increases the ponderomotive force $F_p$ exerted by the laser over the target electrons, since $F_p$ is proportional to the gradient of the laser intensity $F_p \propto -\partial_x \sqrt{1+ p_x^2 +a_0^2} \ $ \cite{Debayle_2017}. The development of an optical shock could also be linked to the steepening of the laser absorption curves of Fig. \ref{fig:ART_all}a (see Ref. \cite{Debayle_2017} and references therein).

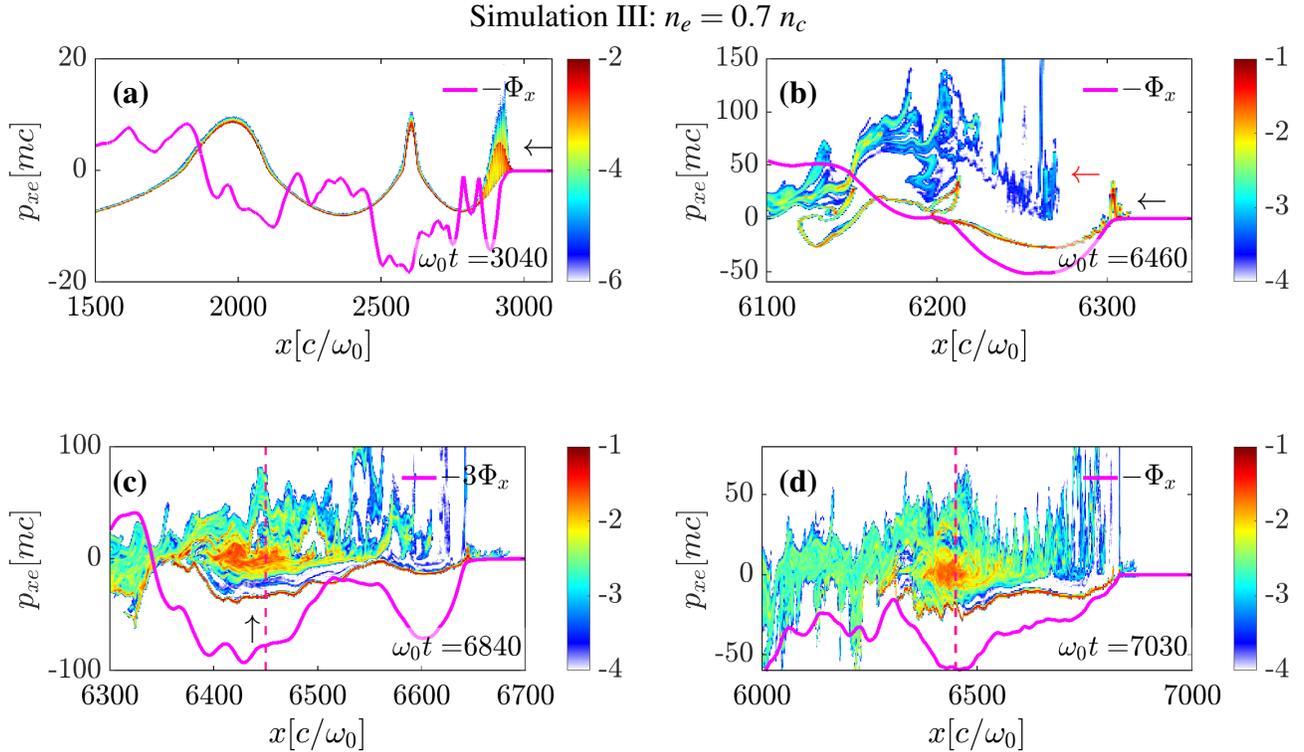
\begin{figure*}[ht!]
	\centering
	\begin{subfigure}[b]{0.49\linewidth}
		\centering
		\includegraphics[width=\linewidth]{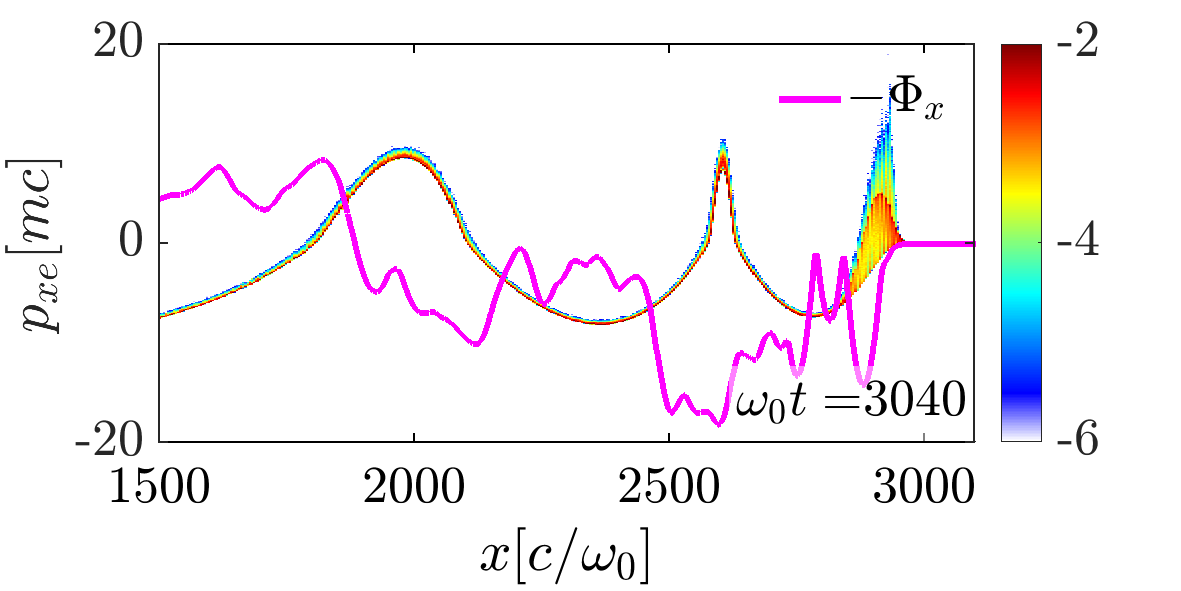}
		\begin{picture}(20,20)
		\put(60,149){\large Simulation \Rnum{3}: $n_{e} = 0.7 \ n_c$}
		\put(-75,120){\large\textbf{(a)}}
		\put(80,100){\large$\leftarrow$}
		\end{picture}	
	\end{subfigure}
	\begin{subfigure}[b]{0.49\linewidth}
		\centering
		\includegraphics[width=\textwidth]{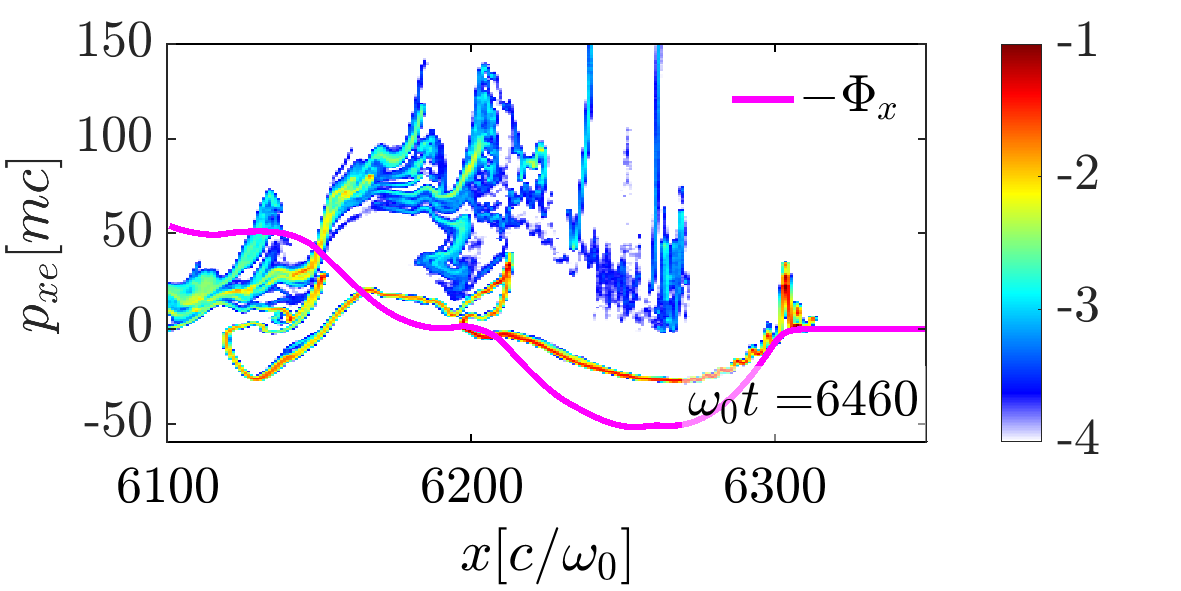}
		\begin{picture}(20,20)
		\put(-75,120){\large\textbf{(b)}}
		\put(35,90){\color{red}\large$\leftarrow$}
		\put(60,80){\large$\leftarrow$}
		\end{picture}
	\end{subfigure}
	\begin{subfigure}[b]{0.49\linewidth}
		\centering
		\includegraphics[width=\linewidth]{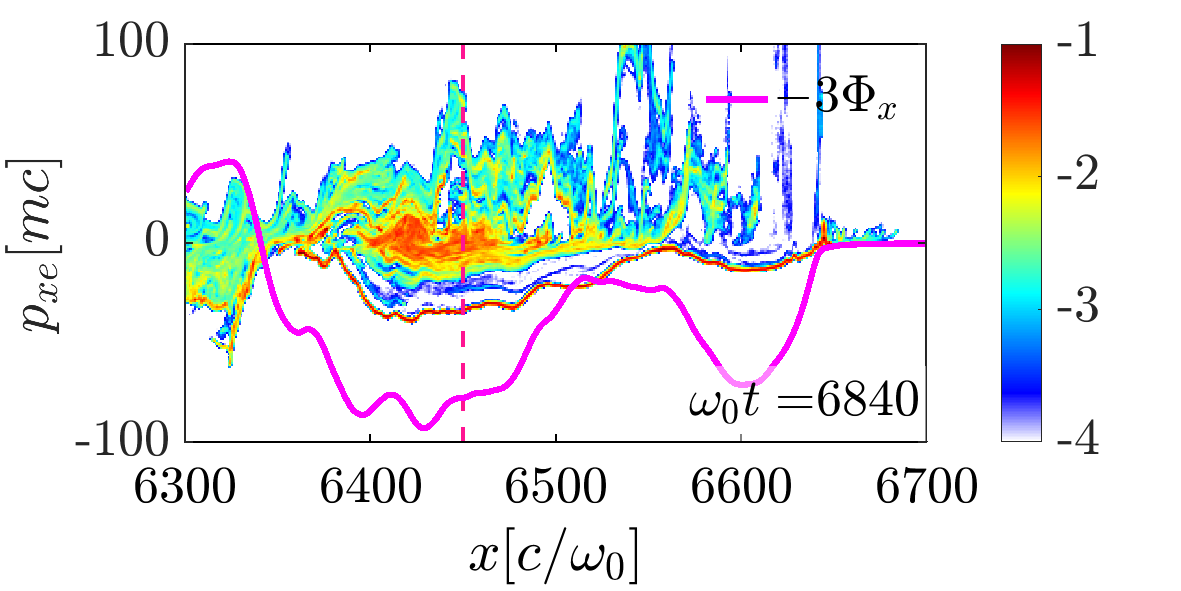}
		\begin{picture}(20,20)
		\put(-75,120){\large\textbf{(c)}}
		\put(-25,65){\large$\uparrow$}
		\end{picture}	
	\end{subfigure}
	\begin{subfigure}[b]{0.49\linewidth}
		\centering
		\includegraphics[width=\linewidth]{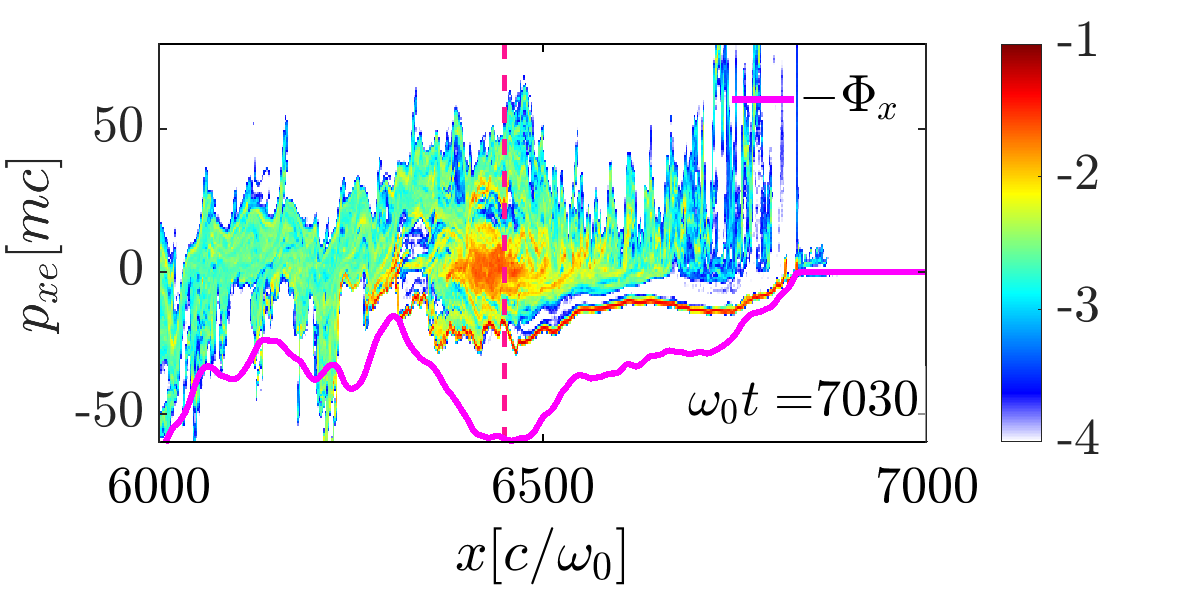}
		\begin{picture}(20,20)
		\put(-75,120){\large\textbf{(d)}}
		\end{picture}	
	\end{subfigure}
	\caption{Electron $(x,p_x)$ phase-space (colorscale) extracted at times $\omega_0t = $ (a) 6270, (b) 6460, (c) 6840 and (d) 7030 corresponding to S.\Rnum{3}. The normalized electrostatic potential $-\Phi_x$ is superposed as solid pink lines.}
	\label{fig:Two_elecron_stream_early}
\end{figure*}

Figures \ref{fig:Two_elecron_stream_early}a-d correspond to the electron $(x,p_x)$ phase spaces of S.\Rnum{3} at times $\omega_0t=$ 3040, 6460, 6840 and 7030. The electrostatic potential $-\Phi_x$ is overlaid as solid pink lines.

Fig. \ref{fig:Two_elecron_stream_early}a shows the early stage of electron energization, that takes place when the laser interacts with the low density wing of the target. The nonlinear wakefield has trapped a bunch of electrons that co-move with it, as indicated by the black arrow in Fig. \ref{fig:Two_elecron_stream_early}a. Note that the electrostatic potential $-\Phi_x$ still comes back to zero in the first oscillations.

Afterwards, at $\omega_0t=$ 6460 (Fig. \ref{fig:Two_elecron_stream_early}b), the laser-plasma interaction gives rise to a two-stream electron distribution resulting from the nonlinear development of the plasma wakefield. The mentioned electron distribution is composed by a cold electron return current and a hot forward-moving one. Note that the laser front has fully steepened by this point, see Fig. \ref{fig:ART_all}d. 

On one hand, the cold electron current contains electrons initially accelerated by the laser wakefield and co-moving with it, see black arrows in Figs. \ref{fig:Two_elecron_stream_early}a and b. The laser-induced displacement of these electrons from their initial position gives rise to a strong charge separation $E_x$ field, which drags them back into the plasma core at a negative momentum $p_x \approx -25$, as seen in Fig \ref{fig:Two_elecron_stream_early}b. This charge separation field, which propagates at a velocity close to the speed of light $c$, is as strong as it is short-lived since the displaced electrons are rapidly pulled back into the plasma. On the other hand, the hot forward electron current is composed of the electrons inside the bulk of the plasma which are already undergoing a heating process in a turbulent regime, as highlighted by the red arrow in Fig \ref{fig:Two_elecron_stream_early}b. At this point, phase mixing between these two electron currents has already started.


At $\omega_0t=$ 6840 (Fig. \ref{fig:Two_elecron_stream_early}c), phase mixing between these high-energy streams continues to develop and results in fast heating of the plasma electrons. The process of breaking of the wakefield and phase mixing between the two cited electron currents is called \textit{beam-loading}.\cite{Debayle_2017} The latter fully develops by $\omega_0t=$ 7030 (Fig. \ref{fig:Two_elecron_stream_early}d). Here the cold return current is re-injected into the target at a negative momentum $p_x \approx -20 \approx -\Phi_x/2$, in agreement with the electron heating model in near-critical plasmas developed in Ref. \cite{Debayle_2017}. The electrostatic potential drops strongly behind the laser pulse and oscillates around $-\Phi = 50$ during the interaction between the two electron currents.

\begin{figure*}[ht!]
	\centering
	\begin{subfigure}[b]{\linewidth}
		\centering
		\includegraphics[width=\linewidth]{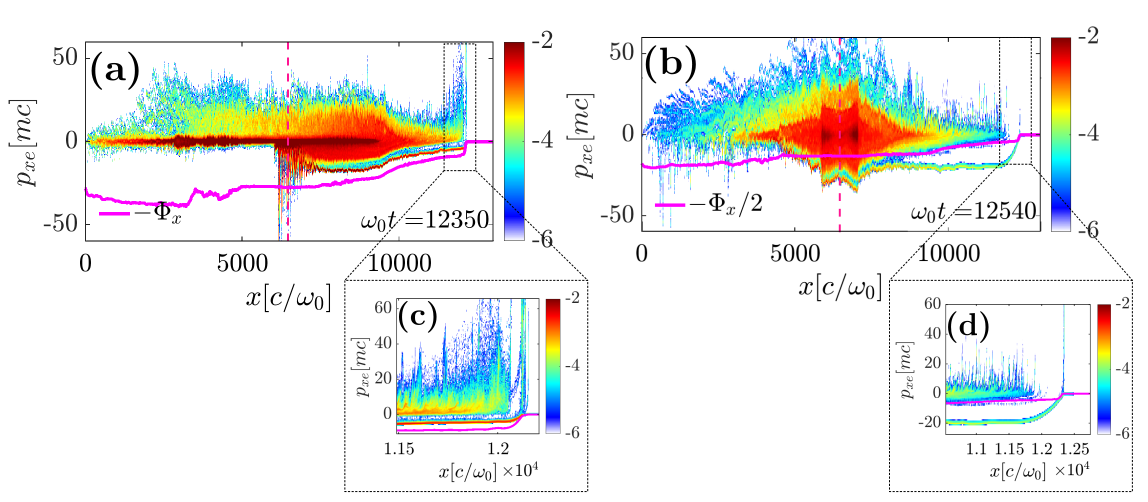}
		\begin{picture}(20,20)
		\put(-45,200){\large$\downarrow$}
		\put(-10,75){\large$\leftarrow$}
		\put(230,75){\large$\leftarrow$}
		\put(-180,230){\large Simulation \Rnum{1}: $n_{e} = 9.8 \ n_c$}
		\put(70,230){\large Simulation \Rnum{3}: $n_{e} = 0.7 \ n_c$}
		\end{picture}	
	\end{subfigure}
	\caption{Late stages electron $(x,p_x)$ phase-spaces (colorscale) corresponding to (a) S.\Rnum{1} and (b) S.\Rnum{3}. The normalized electrostatic potential $-\Phi_x$ is overlaid as solid pink lines.}
	\label{fig:Two_elecron_stream}
\end{figure*}
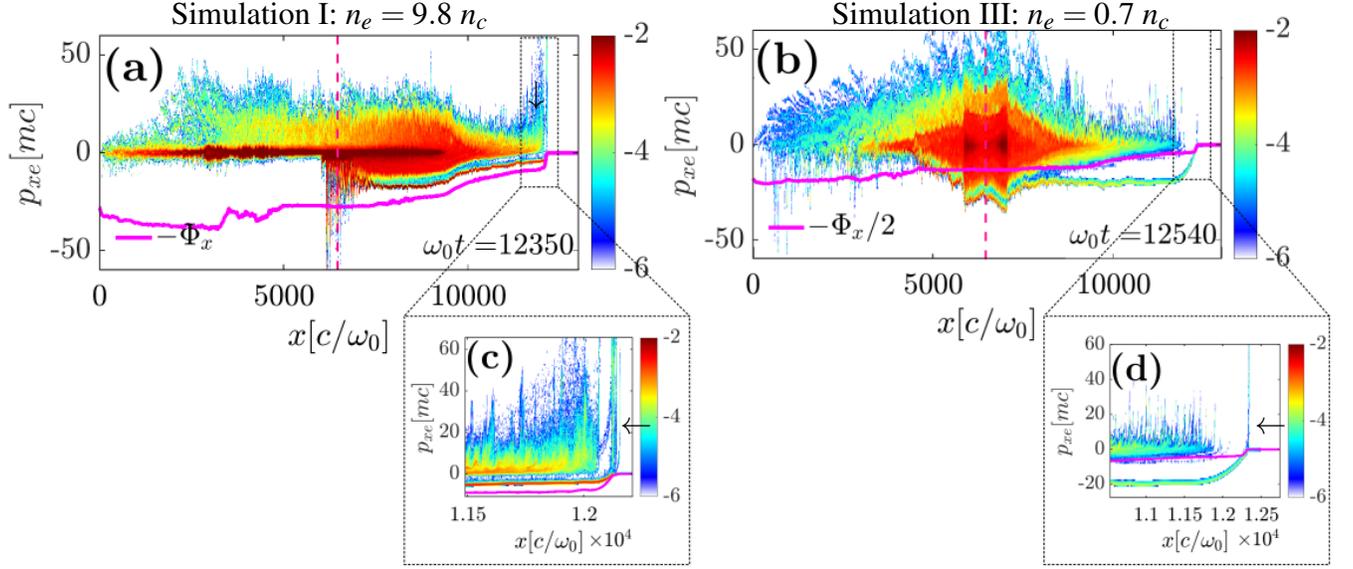

Figures \ref{fig:Two_elecron_stream}a and b correspond to the late stage of the electron energization of Sims. \Rnum{1} and \Rnum{3}, as seen in the electron $(x,p_x)$ phase spaces taken at times $\omega_0t=$ 12350 and 12450, respectively. Figures \ref{fig:Two_elecron_stream}c and d are zooms of the gas down-ramp corresponding to Sims. \Rnum{1} and \Rnum{3}, respectively. The electrostatic potential $-\Phi_x$ is overlaid as a pink solid line in all panels. 

In the high-density S.\Rnum{1} (Figs. \ref{fig:Two_elecron_stream}a and c), the majority of hot electrons are produced at $x \approx 3000 \ c/\omega_0$, when the wakefield resonance condition is satisfied, as previously discussed. Since the laser is completely absorbed at $\omega_0t \approx 4500$, this electron population must heat-up all the gas left to its right. The neutralization of this hot electron population is ensured until the point where it reaches a background density similar to that at which it was created (i.e. at $x \approx 9000 \ c/\omega_0$, see density profiles of Fig. \ref{fig:Density_lineouts}). At this point, charge neutralization can no longer be ensured and a charge separation $E_x$ field is created. This field pulls back the hot electrons, obliging them to recirculate around the target (see black arrow in Fig. \ref{fig:Two_elecron_stream}a), yielding the hotter electron distribution at the right of the density peak. 

In the moderate density S.\Rnum{3} (Figs. \ref{fig:Two_elecron_stream}b and d), the majority of hot electrons are created at $x \approx 5000 \ c/\omega_0$. Hence, charge neutrality is no longer ensured when the hot electrons reach the symmetrical point in the up-ramp, i.e. $x \approx 7800 \ c/\omega_0$. The electrostatic fields arising from the strong electron density and pressure gradients cause the electrons to recirculate back and forth (forming a vortex in the phase space) around the maximum density, while triggering TNSA of the local ions located at the density peak, as will be discussed in the following section.

In S.\Rnum{1} (S.\Rnum{3}) the transmitted percentage of the laser energy is $0.4\%$ ($4.6\%$). The laser can then not be responsible for the electron acceleration seen in the right plasma-vacuum boundary up to momenta $p_x> 60$ in both simulations (see black arrows in Figs. \ref{fig:Two_elecron_stream}c and d). This high-energy electron bunch was initially energized in the plasma wakefield, crossed the entire target and exits now into vacuum. The formation of the return current is seen on the fast-ionizing front of the hot electron streams.

\section{Ion acceleration mechanisms}
\label{Ion_accel}

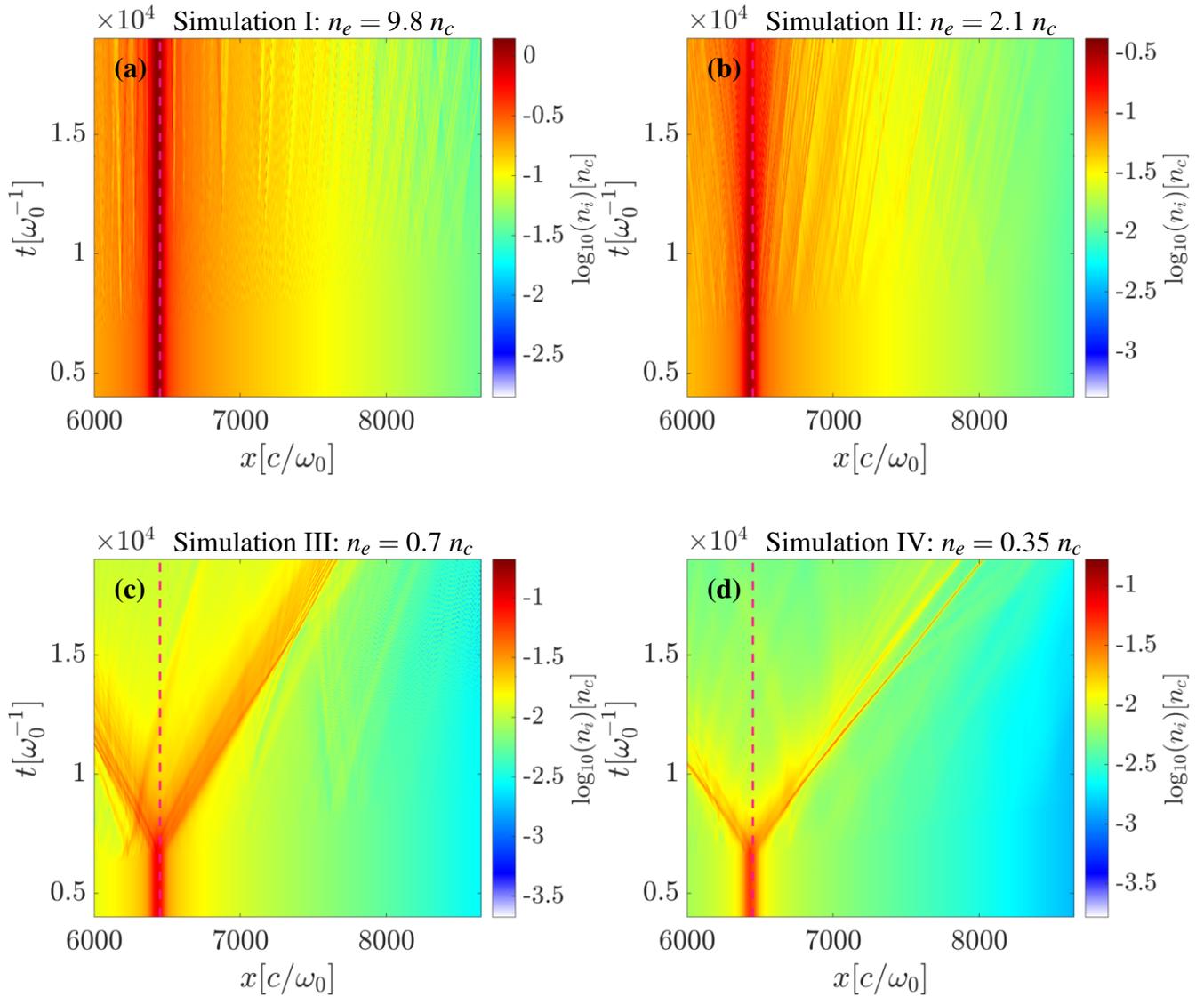
\begin{figure*}[ht!]
	\centering
	\begin{subfigure}[b]{0.49\linewidth}
		\centering
		\includegraphics[width=\linewidth]{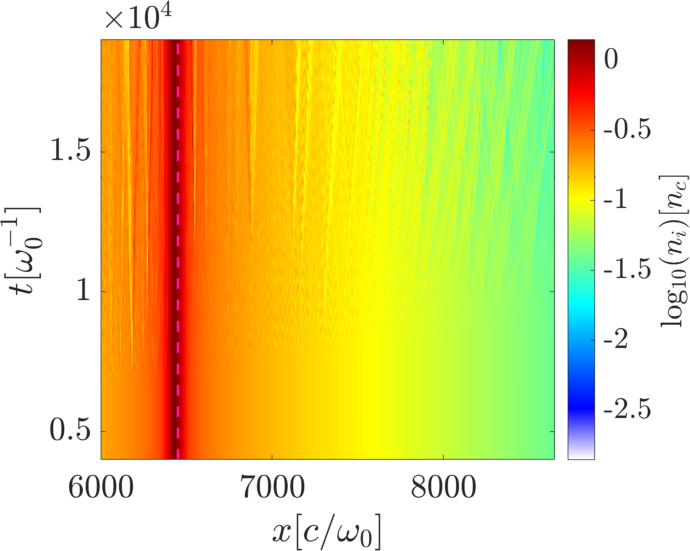}
		\begin{picture}(20,20)
		\put(-45,210){\large Simulation \Rnum{1}: $n_{e} = 9.8 \ n_c$}
		\put(-70,190){\large\textbf{(a)}}
		\end{picture}	
	\end{subfigure}
	\begin{subfigure}[b]{0.49\linewidth}
		\centering
		\includegraphics[width=\textwidth]{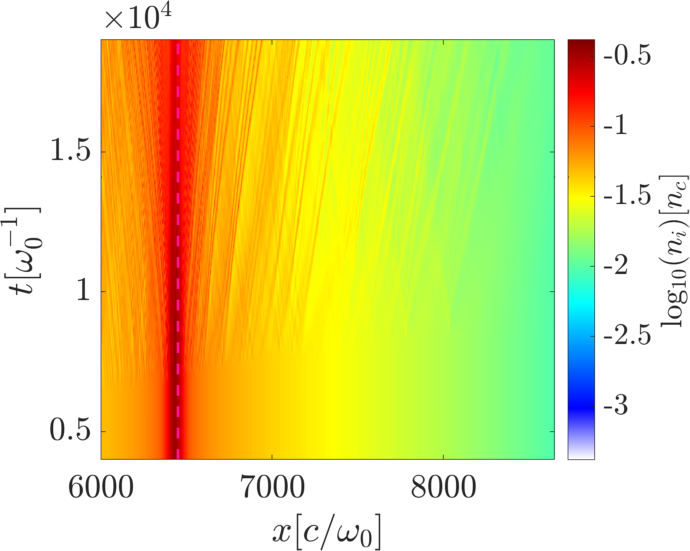}
		\begin{picture}(20,20)
		\put(-45,210){\large Simulation \Rnum{2}: $n_{e} = 2.1 \ n_c$}
		\put(-70,190){\large\textbf{(b)}}
		\end{picture}
	\end{subfigure}
	\begin{subfigure}[b]{0.49\linewidth}
		\centering
		\includegraphics[width=\linewidth]{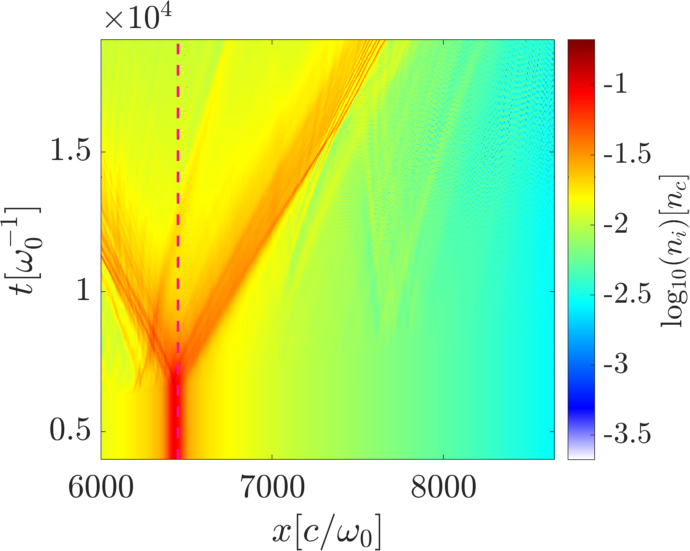}
		\label{fig:Two_stream_c}
		\begin{picture}(20,20)
		\put(-45,210){\large Simulation \Rnum{3}: $n_{e} = 0.7 \ n_c$}
		\put(-70,190){\large\textbf{(c)}}
		\end{picture}	
	\end{subfigure}
	\begin{subfigure}[b]{0.49\linewidth}
		\centering
		\includegraphics[width=\linewidth]{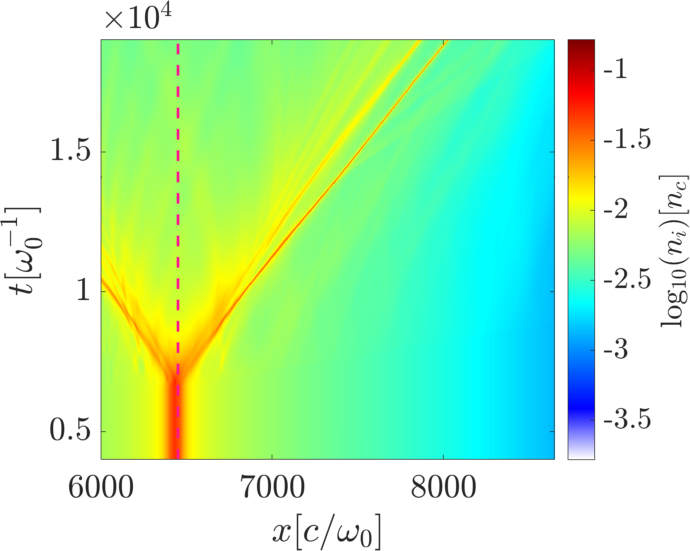}
		\label{fig:Two_stream_d}
		\begin{picture}(20,20)
		\put(-45,210){\large Simulation \Rnum{4}: $n_{e} = 0.35 \ n_c$}
		\put(-70,190){\large\textbf{(d)}}
		\end{picture}	
	\end{subfigure}
	\caption{$(x,t)$ maps of the $\mathrm{N^{7+}}$ ion density centered on the target down-ramp for the (a) Simulation \Rnum{1} with $n_{at,max} = 1.4 \ n_c$, (b) Simulation \Rnum{2} with $n_{at,max} = 0.3 \ n_c$, (c) Simulation \Rnum{3} with $n_{at,max} = 0.1 \ n_c$ and (d) Simulation \Rnum{4} with $n_{at,max} = 0.05 \ n_c$, see Table \ref{tab:summary_simulations}. The density peak is marked with a pink dashed line.}
	\label{fig:ion_density_charts}
\end{figure*}

The subsequent ion acceleration mechanisms and their efficiency chiefly rely on the laser absorption conditions. Three regimes of ion acceleration, illustrated in the $(x,t)$ ion density maps of Figs. \ref{fig:ion_density_charts}a-d (corresponding to Sims. \Rnum{1} - \Rnum{4}), stem from the 1-D simulations. 

In the high density cases corresponding to Sims. \Rnum{1} and \Rnum{2} (Figs. \ref{fig:ion_density_charts}a and b), the laser is absorbed in the gas up-ramp. The hot electrons propagate through the gas and trigger ion acoustic waves (IAWs) propagating at $\pm\,C_s$ with equal amplitudes. Here $C_s=\sqrt{Z^*T_e/m_i} \approx 0.025 \ c$ is the local ion acoustic speed (with $Z^* \approx 5$ and $T_e \approx 3$, the ionic charge and electron temperature, respectively). 

For the moderate gas densities of Sims. \Rnum{3} and \Rnum{4} (Figs. \ref{fig:ion_density_charts}c and d), the laser drives strong wakefields that trap a large number of electrons, rapidly turn turbulent and heat the plasma. The strong electron pressure triggers an electrostatic shock as illustrated in Fig. \ref{fig:ion_density_charts}c and d. Background ions reflection takes place at $t \gg \omega_{pi}^{-1}$. In S.\Rnum{3} (S.\Rnum{4}) the front of the expanding plasma travels at $v_{i,w} \approx 0.08 \ c, M \approx 2$ ($v_{i,w} \approx 0.13 \ c, M \approx 2$), where $M=v_{x,i}/C_s$ is the upstream Mach number. As one can deduce, $T_e$ is higher in S.\Rnum{4} than in S.\Rnum{3}. These cases correspond to the ideal laser absorption regime and are discussed in the following Section \ref{CSA}.

Finally, for low gas densities (S.\Rnum{5}), the laser absorption essentially remains in a moderately non-linear wakefield regime. The electron temperature is not strong enough to trigger a shock and only TNSA subsists. 

\subsection{Ion acceleration in moderate density simulations: collisionless electrostatic shock formation}
\label{CSA}

\begin{figure*}[ht!]
	\centering
	\begin{subfigure}[b]{0.49\linewidth}
		\centering
		\includegraphics[width=\linewidth]{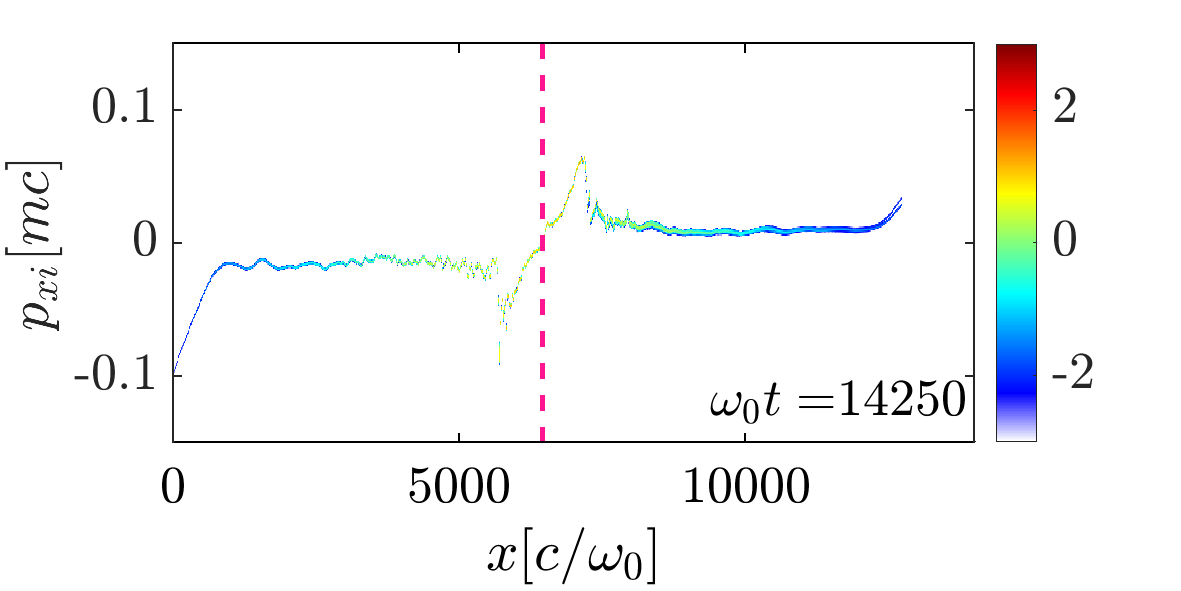}
		\begin{picture}(20,20)
		\put(-50,142){\large Simulation \Rnum{3}: $n_{e} = 0.7 \ n_c$}
		\put(-70,120){\large\textbf{(a)}}
		\end{picture}	
	\end{subfigure}
	\begin{subfigure}[b]{0.49\linewidth}
		\centering
		\includegraphics[width=\textwidth]{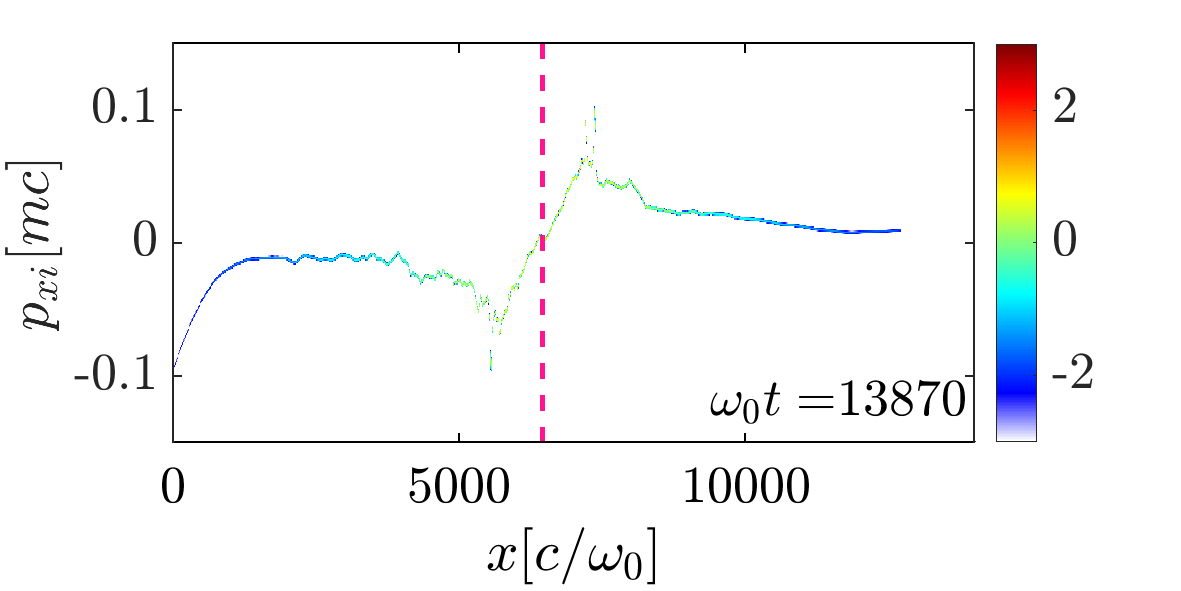}
		\begin{picture}(20,20)
		\put(-50,142){\large Simulation \Rnum{4}: $n_{e} = 0.35 \ n_c$}
		\put(-70,120){\large\textbf{(b)}}
		\end{picture}
	\end{subfigure}
	\caption{{\color{black}Ion phase-spaces for the (a) simulation \Rnum{3} with $n_{at,max} = 0.1 \ n_c$ and (b) simulation \Rnum{4} with $n_{at,max} = 0.05 \ n_c$ taken at the shock formation instants when background ion reflection is about to start.}}
	\label{fig:shock_formation}
\end{figure*}
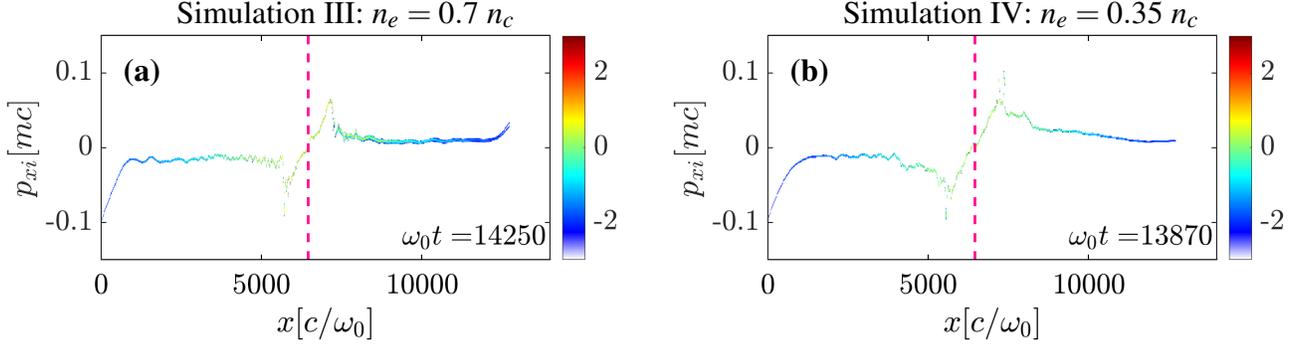

Figures \ref{fig:shock_formation}a and b show the ion $(x,p_x)$ phase spaces for the moderate density Sims. \Rnum{3} and \Rnum{4} at the shock formation time, i.e. the onset of ion reflection. At this point, the ion velocity profile has fully steepened. In S.\Rnum{3} shock reflection starts at $\omega_0t \approx 14250$ ($7.6\,\mathrm{ps}$), and in the lower density S.\Rnum{4} it does so at $\omega_0t \approx 13870$ ($7.4\,\mathrm{ps}$). The shock formation time does not satisfy the normally assumed proportionality $t \propto \omega_{pi}^{-1}$. Shock reflection could be accelerated in S.\Rnum{4}, in respect to S.\Rnum{3}, by widening the density profile until achieving optimum plasma heating conditions, i.e. the transition from a laminar to a turbulent electron energizing regime while crossing the gas density peak (as in S.\Rnum{3}, see absorption curves of Fig. \ref{fig:ART_all}). 
\\

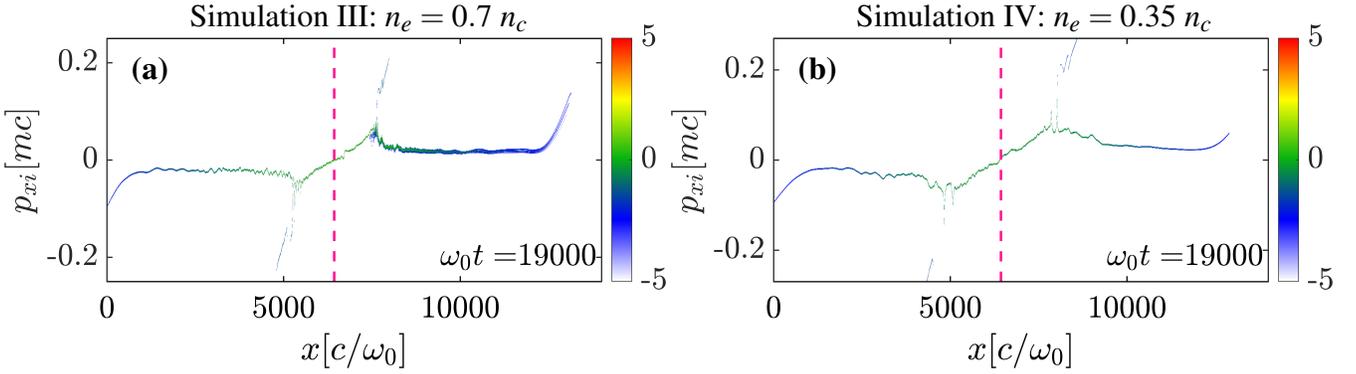
\begin{figure*}[ht!]
	\centering
	\begin{subfigure}[b]{0.49\linewidth}
		\centering
		\includegraphics[width=\linewidth]{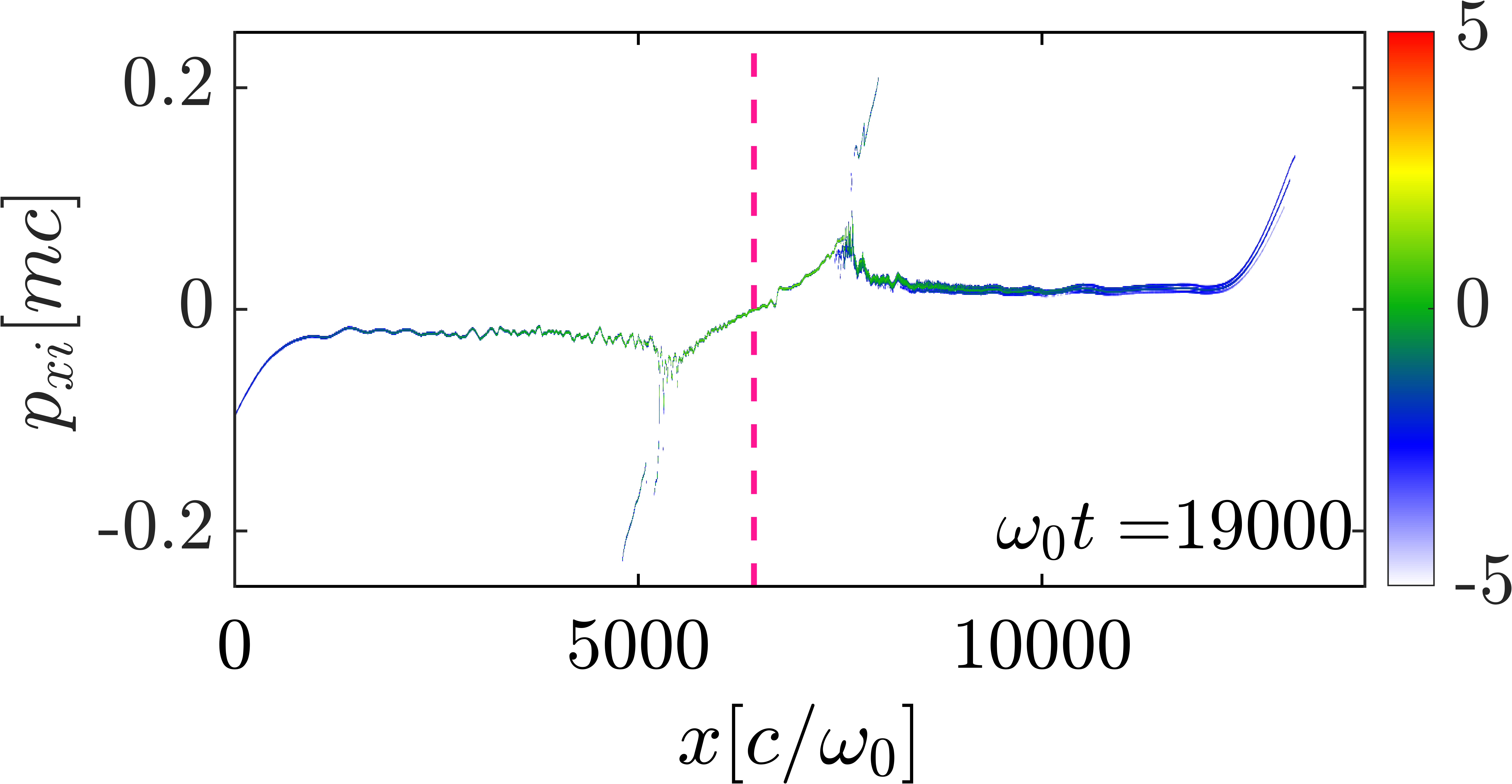}
		\begin{picture}(20,20)
		\put(-45,150){\large Simulation \Rnum{3}: $n_{e} = 0.7 \ n_c$}
		\put(-67,130){\large\textbf{(a)}}
		\end{picture}	
	\end{subfigure}
	\begin{subfigure}[b]{0.49\linewidth}
		\centering
		\includegraphics[width=\textwidth]{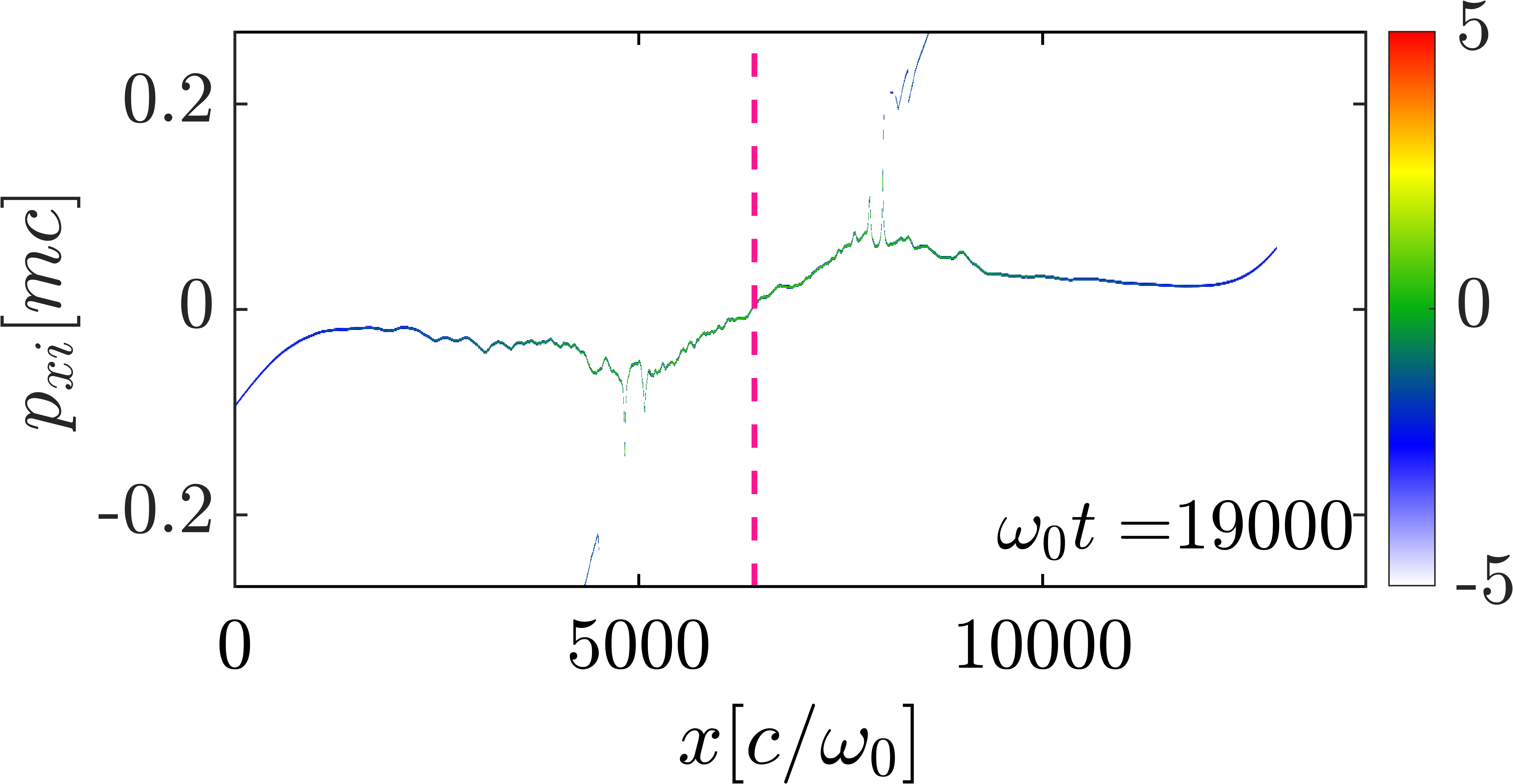}
		\begin{picture}(20,20)
		\put(-45,150){\large Simulation \Rnum{4}: $n_{e} = 0.35 \ n_c$}
		\put(-67,130){\large\textbf{(b)}}
		\end{picture}
	\end{subfigure}
	\caption{Final $\mathrm{N^{7+}}$ ion $(x,p_x)$ phase spaces of the (a) $n_{at,max}$ = 0.1 $n_c$ and (b) $n_{at,max}$ = 0.05 $n_c$ maximum density simulations.}
	\label{fig:Final_ion}
\end{figure*}

Figures \ref{fig:Final_ion}a and b show the ion $(x,p_x)$ phase spaces corresponding to the last time of Sims. \Rnum{3} and \Rnum{4}. CES formation was observed in both simulations at these times. 
\\

In these cases, the accelerated ion velocity profile continually steepens when moving down the density gradients while the quasi-exponential density "wings" are subject to a uniform slower acceleration in the ambipolar field $E_x \approx T_e/L_n$, where $L_n$ is the quasi-constant density gradient. This process bears resemblance to wavebreaking in TNSA within a pre-expanded plasma profile,\cite{Grismayer_2006} and has been described as the expansion of a hot plasma into a cold rarefied plasma.\cite{Perego_2013,Nechaev_2020} The steepening of the ion velocity profile goes along with an increase in the local potential barrier up to the point of reflecting the upstream ions, uniformly accelerated by the ambipolar field. An ion located in front of the shock and moving at a velocity $v_0 - v_s$ in the shock frame, where $v_s$ is the lab-frame shock front velocity and $v_0$ is the initial lab-frame ion velocity, gets reflected from the shock front if its kinetic energy in the shock frame, $E_k$, is lower than the electrostatic potential $Z^*e\Delta \phi$ experienced by the particle in the shock region:

\begin{eqnarray}
\label{eq:shock_cond}
E_k = \frac{m_i(v_0-v_s)^2}{2} \ < \ Z^*e\Delta \phi.
\end{eqnarray}

\noindent
Here, $m_i$ is the ion mass and $Z^*$ the ionization degree. Shock formation was observed in the simulations with peak electron density between 0.35 $n_c$ and 0.7 $n_c$ (Sims. \Rnum{3} and \Rnum{4}). Moreover, the uniform slower acceleration of the ions located in the target wings is due to the constant plasma scale length $L_n$ of the density profile, as seen in Fig. \ref{fig:Final_ion}, which gives rise to a constant TNSA derived electrostatic field. The contrary would induce a chirp in the reflected ions, making them lose their CSA characteristic peaked spectra.\cite{Fiuza_2012} The CES formed in Sims. \Rnum{3} and \Rnum{4} have Mach numbers $M \approx $1.6 and $M \approx $ 2.4, respectively. Mach numbers between $1.6 \leq M \leq 3$ are characteristic of super-critical shocks, in which ion reflection and plasma heating are the stabilizing energy dissipation mechanisms.\cite{Fiuza_2012} The obtained Mach number values agree with the CES conditions in near-critical plasmas studied by F. Fiuza \textit{et al.,} \cite{Fiuza_2013,Fiuza_2012} and M.E. Dieckmann \textit{et al.,}.\cite{Dieckmann_2013a,Dieckmann_2013b} 

\begin{table*}[ht!]
	\caption{\label{tab:shock_reflection__orig} Plasma parameters relative to the shock reflection condition in S.\Rnum{3} ($n_{e} = 0.7 \ n_c$) for the $\mathrm{N^{7+}}$ and $\mathrm{N^{6+}}$ ion species.}
		\begin{tabular}{|c|c|c|c|c|c|c|c|c|c|c|c|}
		\hline
			Ion & Time [$\omega_0^{-1}$] & $m_i$ [$m_e$] & $v_f$ [$c$] & $v_0$ [$c$]  & $(v_f - v_0)$ [$c$] & $E_k$  [$m_ec^2$]& $Z^*$ &$ \Delta\phi$ [$m_ec^2/e$] &$ Z^*\Delta\phi$ [$m_ec^2/e$] & $C_s$ [$c$] & M\\
			\hline
			$\mathrm{N^{7+}}$      & 13945            & 25704 & 0.109 & 0.030 & 0.079 & 79.804 & 7 & 12 & 84 & 0.048 & 1.642 \\
			$\mathrm{N^{6+}}$       &  13945          &  25704 & 0.109 & 0.020 & 0.089 & 101.34 & 6 & 12 & 72 & 0.048 & 1.850 \\
			\hline
		\end{tabular}
\end{table*}

The plasma parameters relevant for shock reflection in S.\Rnum{3} are summarized in Table \ref{tab:shock_reflection__orig}. The shock reflection condition, corresponding to Eq. \ref{eq:shock_cond}, is fulfilled for the $\mathrm{N^{7+}}$ ionic species at $\omega_0t = 13495$. In the case of $\mathrm{N^{6+}}$ ions the kinetic energy exceeds the electrostatic potential experienced by the ionic species in the shock region $E_k > Z^*e\Delta\phi$, thus preventing electrostatic reflection, as discussed hereafter. 

\subsubsection{Effect of the ionization degree on shock reflection}

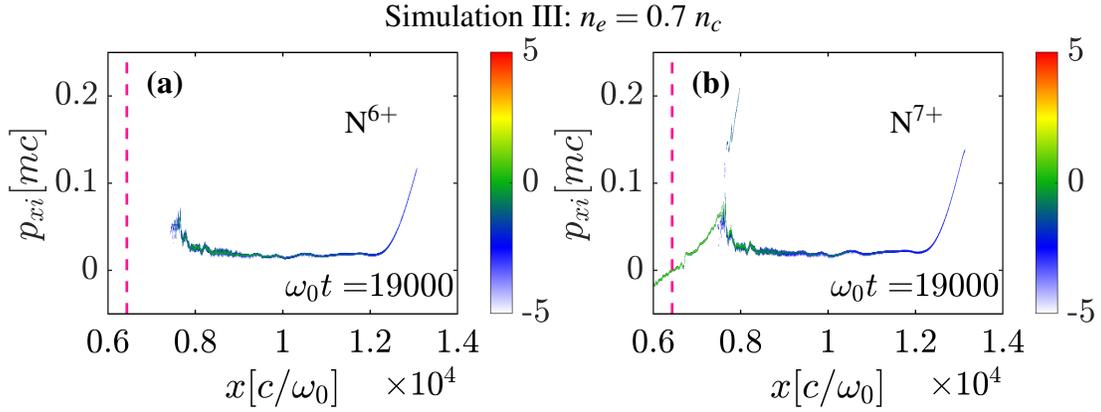
\begin{figure*}
	\centering
	\begin{subfigure}[b]{0.40\linewidth}
		\centering
		\includegraphics[width=\textwidth]{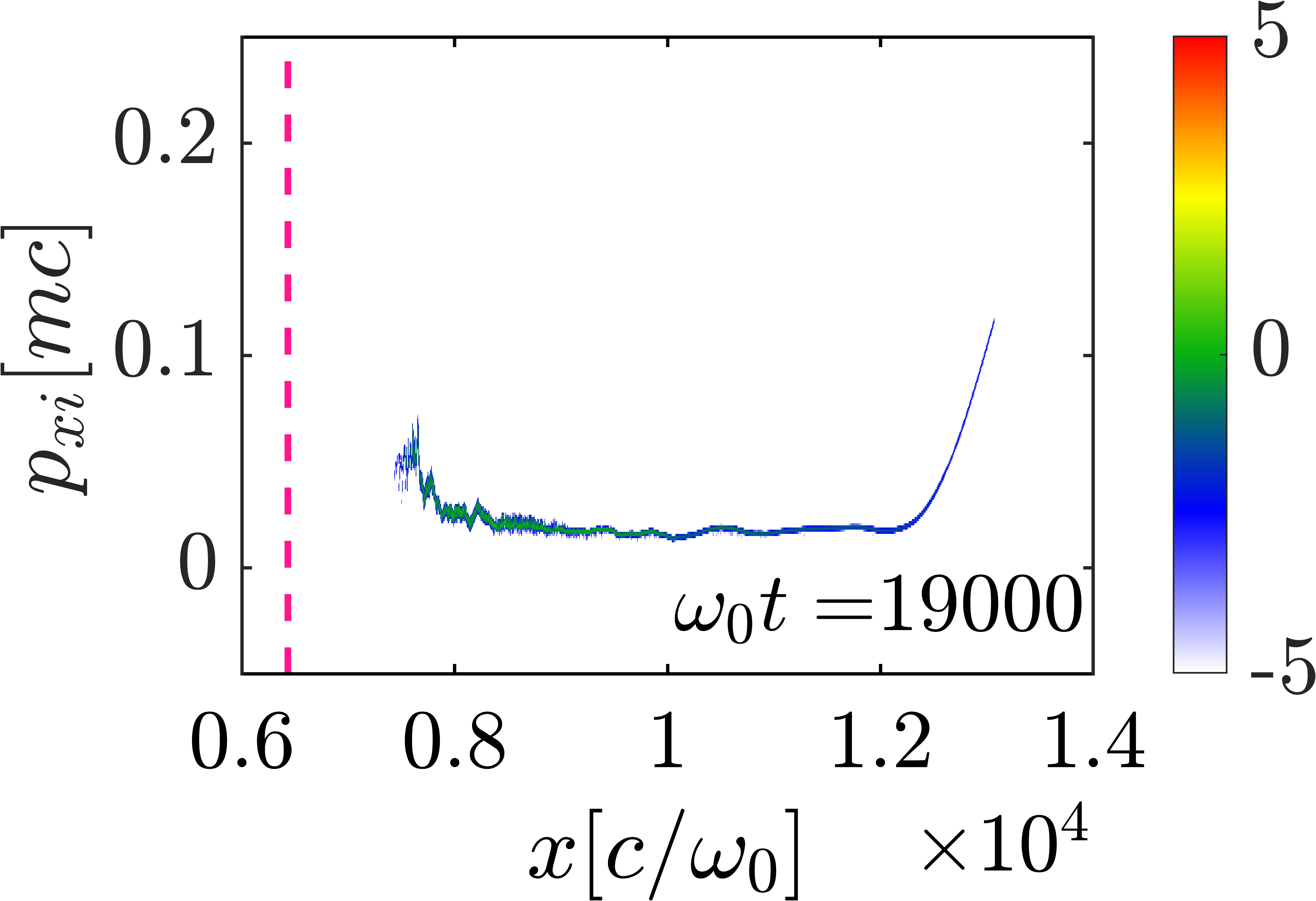}
		\begin{picture}(20,20)
		\put(50,165){\large Simulation \Rnum{3}: $n_{e} = 0.7 \ n_c$}
		\put(-40,140){\large\textbf{(a)}}
		\put(35,125){\large\textbf{$\mathrm{N^{6+}}$}}
		\end{picture}
	\end{subfigure}
	\begin{subfigure}[b]{0.40\linewidth}
		\centering
		\includegraphics[width=\textwidth]{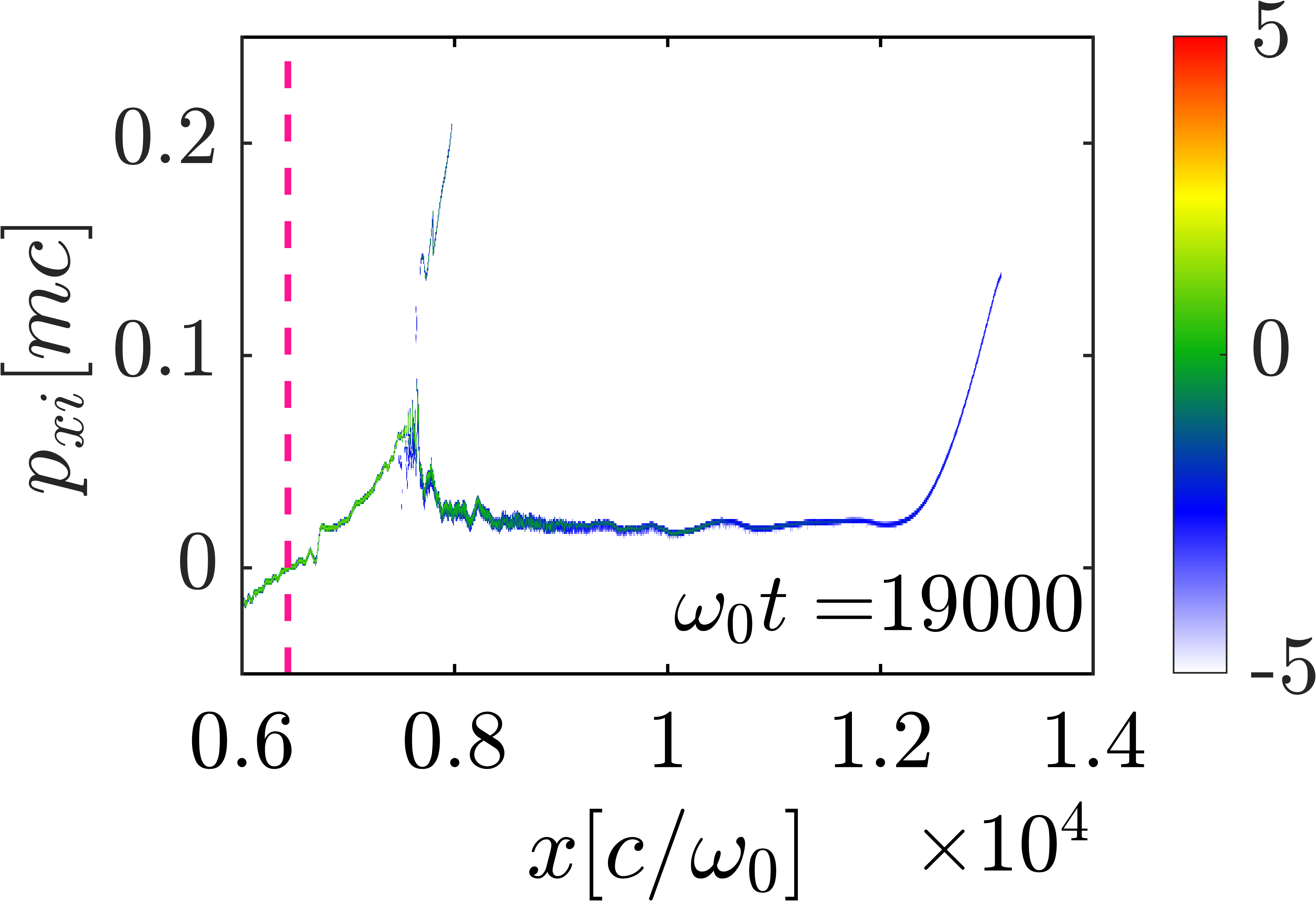}
		\begin{picture}(20,20)
		\put(-40,140){\large\textbf{(b)}}
		\put(35,125){\large\textbf{$\mathrm{N^{7+}}$}}
		\end{picture}
	\end{subfigure}
	\caption{Influence of the ionization state in shock reflection as seen in the ion phase spaces of the (a) $\mathrm{N^{6+}}$ and (b) $\mathrm{N^{7+}}$ ion charged species phase spaces corresponding to S.\Rnum{3}.}
	\label{fig:Ioniz_state}
\end{figure*}

In S.\Rnum{3} the gas down-ramp has experienced a strongly attenuated laser pulse, and therefore has not been fully ionized. Hence, it is interesting to examine how ionic species of different charge state interact with a CES and, in particular, whether they can be reflected by it. Recalling the shock reflection condition of Eq. \ref{eq:shock_cond}, the latter is highly sensitive to the ionic species ionization state, since it will determine the electrostatic potential energy experienced by the incoming ions. Figs. \ref{fig:Ioniz_state}a and b focus on the ion $(x,p_x)$ phase space down-ramp of S.\Rnum{3}, corresponding to the $\mathrm{N^{6+}}$ and $\mathrm{N^{7+}}$ nitrogen ionic species, respectively. The $\mathrm{N^{6+}}$ ion species do not see a sufficiently strong electrostatic potential and are consequently not reflected from the shock front, located at $x = 7500 \ c/\omega_0$ in Fig. \ref{fig:Ioniz_state}a. Instead, they are slightly accelerated and finally cross the shock region. Only the fully ionized $\mathrm{N^{7+}}$ ions experience the maximum electrostatic potential and are reflected at twice the shock velocity, see Fig. \ref{fig:Ioniz_state}b.

\subsubsection{Role of laser-driven ion acceleration combined with the gas profile in shock formation}
\label{laser_free}

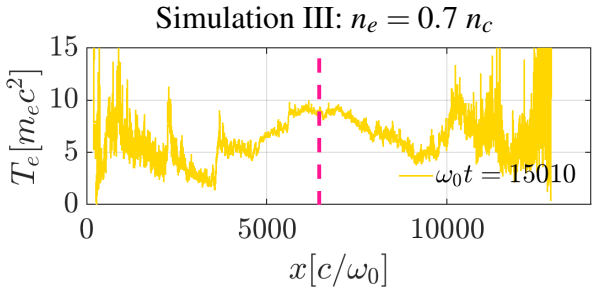
\begin{figure}[ht!]
	\centering
	\begin{subfigure}[b]{\columnwidth}
		\centering
		\includegraphics[width=\linewidth]{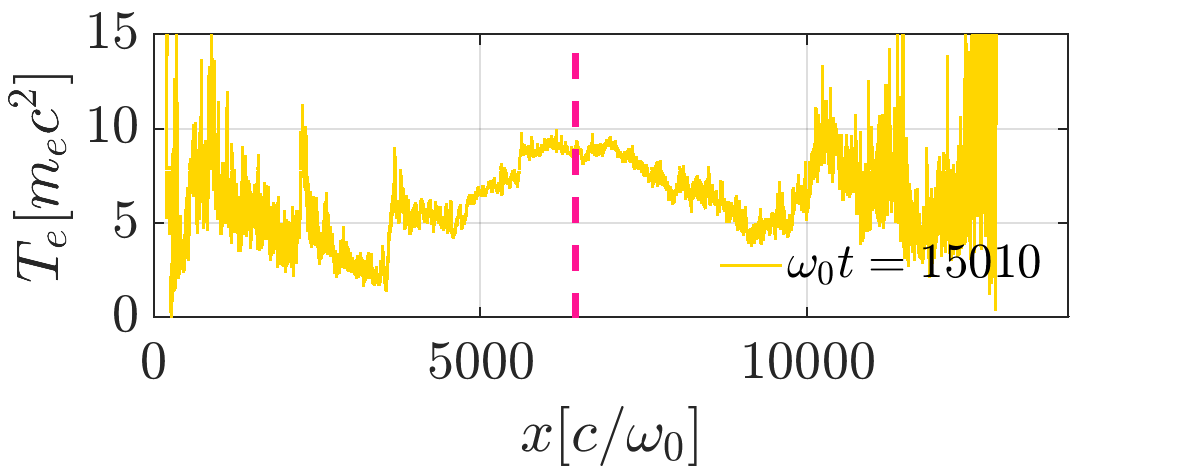}
		\begin{picture}(20,20)
		\put(-55,120){\large Simulation \Rnum{3}: $n_{e} = 0.7 \ n_c$}
		\end{picture}	
	\end{subfigure}
	\vspace{-20pt}
	\caption{Electron temperature $T_e$ lineout extracted from S.\Rnum{3} at $\omega_0t = 15010$.}
	\label{fig:Te_S3}
\end{figure}

So far, we have discussed the formation of CES in terms of the slow plasma expansion driven by the uniformly heated bulk electrons, leaving aside the possible influence of shorter-scale velocity of ion density perturbations imparted by the laser. The direct laser impact on ion acceleration and shock formation was assessed by performing a laser-free simulation with a pre-heated electron population with $T_e = a_0  = 11$ and $n_{e,max} = 0.7 \ n_c$ (S.\Rnum{6}). The chosen $T_e = a_0$ value approximately agrees with the mean electronic temperature found in the original laser-on S.\Rnum{3}, as seen in Fig. \ref{fig:Te_S3}.
\\

\begin{figure*}[ht!]
	\centering
	\begin{subfigure}[b]{0.49\linewidth}
		\centering
		\includegraphics[width=\linewidth]{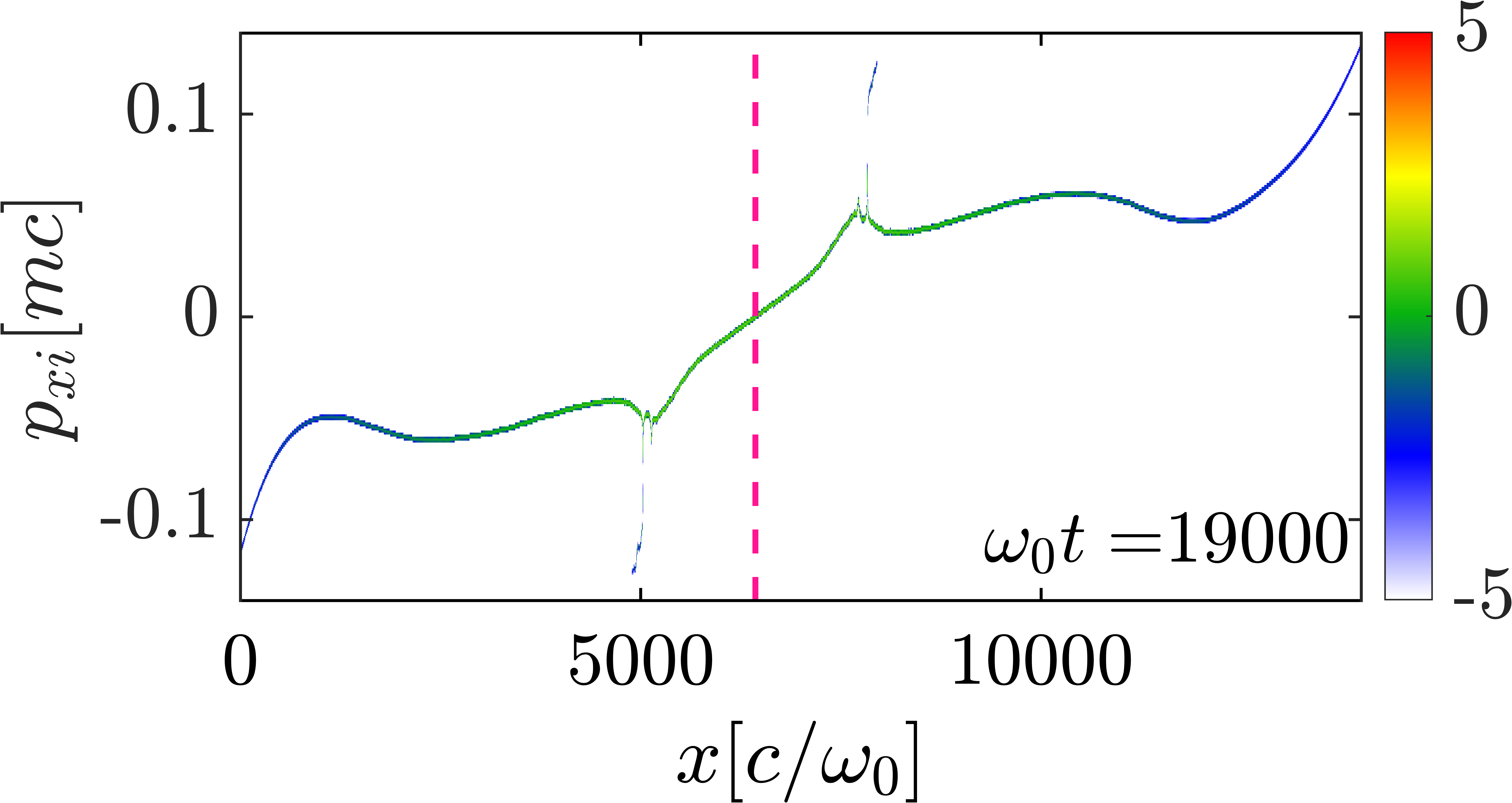}
		\begin{picture}(20,20)
		\put(-45,155){\large Simulation \Rnum{6}: laser-free}
		\put(-67,135){\large\textbf{(a)}}
		\end{picture}	
	\end{subfigure}
	\begin{subfigure}[b]{0.49\linewidth}
		\centering
		\includegraphics[width=\textwidth]{./Images/new_images/Ion_accel_b_ion.png}
		\begin{picture}(20,20)
		\put(-45,155){\large Simulation \Rnum{3}: laser-on}
		\put(-67,135){\large\textbf{(b)}}
		\end{picture}
	\end{subfigure}
	\begin{subfigure}[b]{0.49\linewidth}
		\centering
		\includegraphics[width=\linewidth]{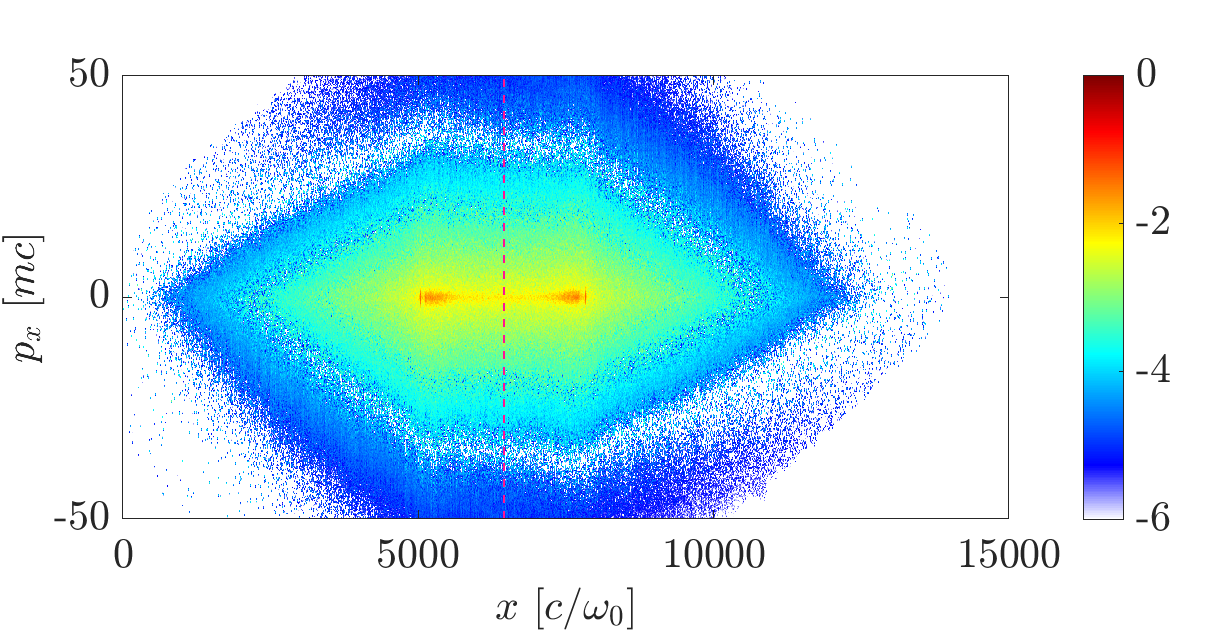}
		\label{fig:Shock_barriers_a}
		\begin{picture}(20,20)
		\put(-80,125){\large\textbf{(c)}}
		\end{picture}	
	\end{subfigure}
	\begin{subfigure}[b]{0.49\linewidth}
		\centering
		\includegraphics[width=\textwidth]{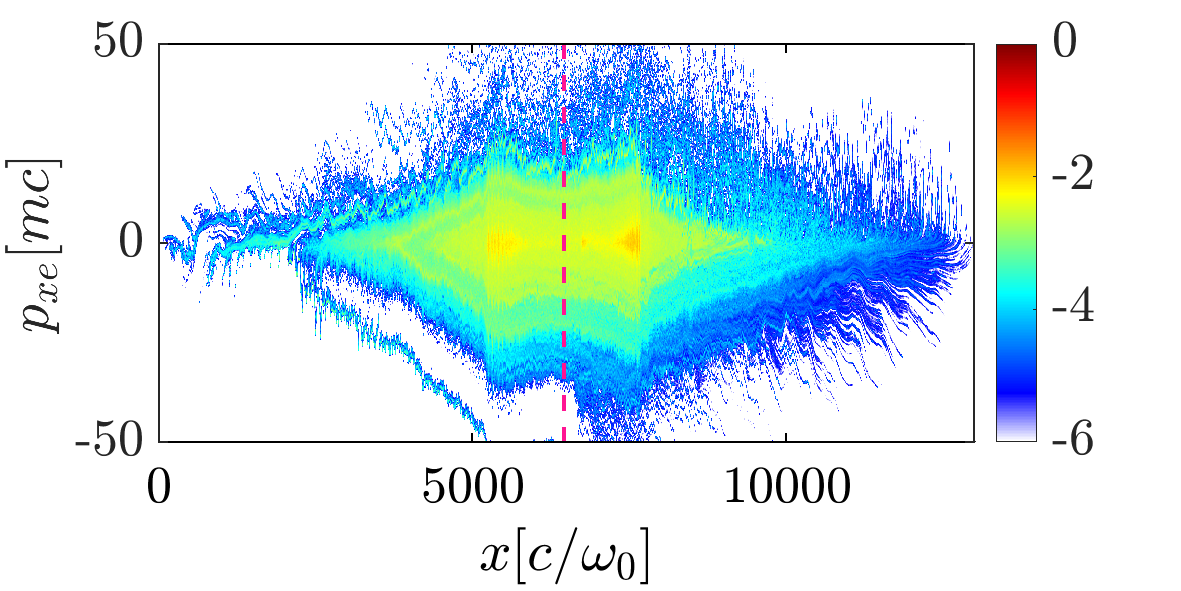}
		\label{fig:Shock_barriers_b}
		\begin{picture}(20,20)
		\put(-80,125){\large\textbf{(d)}}
		\end{picture}
	\end{subfigure}
	\caption{(a) Final ion $(x,p_x)$ phase spaces corresponding to (a) S.\Rnum{6} and (b) S.\Rnum{3}. Electrostatic shock barriers in (c) S.\Rnum{6} and (d) S.\Rnum{3}, as seen in the electron $(x,p_x)$ phase spaces of the respective simulations.}
	\label{fig:Final_ion_a0_0}
\end{figure*}

Figures \ref{fig:Final_ion_a0_0}a and b show the ion $(x,p_x)$ phase spaces corresponding to the last time step of S.\Rnum{6} (with a pre-heated Maxwellian electron population) and S.\Rnum{3}, respectively. Figures \ref{fig:Final_ion_a0_0}c and d plot the corresponding electron $(x,p_x)$ phase spaces. The longitudinal electrostatic field $E_x$ is overlaid as dark violet lines. The laser-free (LF) simulation (S.\Rnum{6}) produces a shock that is qualitatively similar to the one triggered in the laser-on (LO) simulation (S.\Rnum{3}), as seen by comparing the ion $(x, p_x)$ phase spaces plotted in Figs. \ref{fig:Final_ion_a0_0}a and b. However, the initial plasma modulations are absent in the laser-free simulation, which confirms their laser-driven nature. 

Forcing a uniform electron temperature through the entire profile tends to enhance TNSA in the profile's wings. As a result, in the quasi exponentially decreasing wings that are subject to a uniform accelerating field, the ion velocity reaches the values $v_i \approx 0.05 \ c$ in the LF simulation  (Fig. \ref{fig:Final_ion_a0_0}a). Note that in the LO simulation $v_i \approx 0.02 \ c$ (Fig. \ref{fig:Final_ion_a0_0}b). This makes it harder for the fast ions accelerated from the density spike to catch up with and eventually reflect the ions accelerated at the gas wings in the LF simulation.

The LO simulation electron $(x,p_x)$ phase space is strongly modulated as a consequence of the direct-laser and wakefield-induced electron acceleration \cite{Debayle_2017,Robinson_2011} (Fig. \ref{fig:Final_ion_a0_0}d), in comparison with the LF simulation (Fig. \ref{fig:Final_ion_a0_0}c). Furthermore, the LO electron phase space shows well defined electrostatic shock fronts moving both forward and backward, as seen in Fig. \ref{fig:Final_ion_a0_0}d at $x = 5000 c/\omega_0$ and $x = 7000 c/\omega_0$. In the LF electron $(x,p_x)$ phase space the electrostatic confinement in the downstream region is weaker. As a result, two diluted electron populations centered at $x = 5000 c/\omega_0$ and $x = 8000 c/\omega_0$ travel both forward and backward down the density gradients, as seen in Fig. \ref{fig:Final_ion_a0_0}c. 

\begin{figure}
	\centering
	\begin{picture}(20,20)
	\put(-50,5){\large Simulation \Rnum{10}: $T_e = a_0/2$}
	\end{picture}
	\includegraphics[width=\columnwidth]{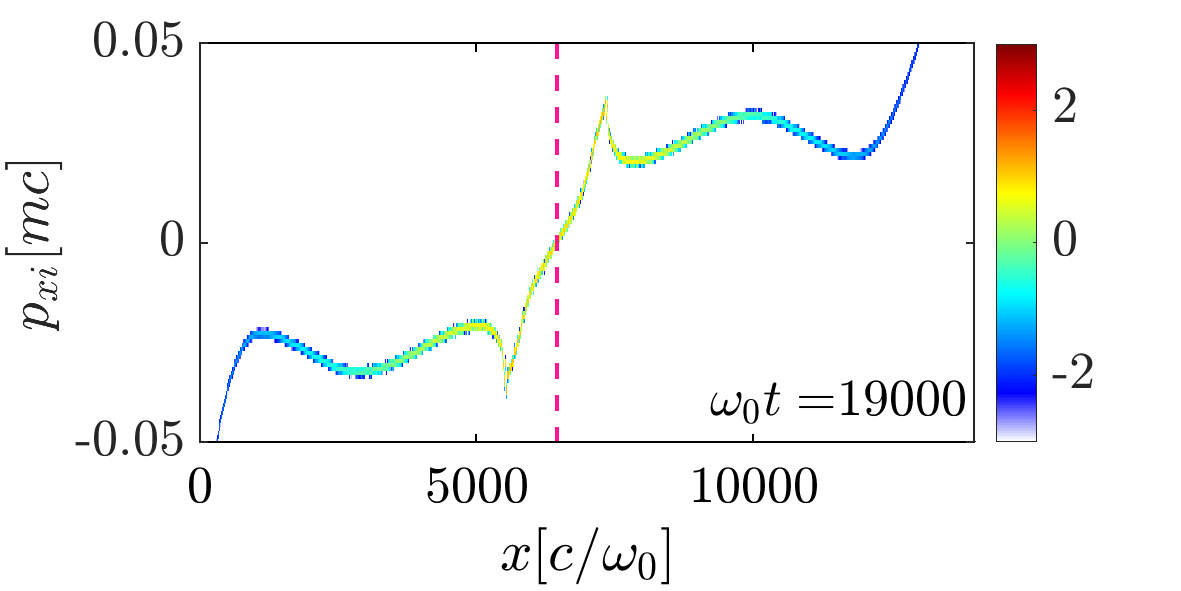}
	\caption{Ion $(x, p_x)$ phase space corresponding to the laser-free Simulation \Rnum{10} with $n_{at,max} = 0.1 \ n_c$ and an initialized hot electron population with $T_e = a_0/2$.}
	\label{fig:Te_a0_over_2}
\end{figure}

A second LF simulation, with an initialized hot electron population with $T_e = a_0/2$ (S.\Rnum{10}), was performed to investigate whether a weaker target energization could trigger an electrostatic shock in less time. We intended to compare the shock formation time with the value obtained in the LF S.\Rnum{6} initialized with $T_e = a_0$, where a shock was triggered at $\omega_0t = 15171$. However, by the end of S.\Rnum{10} no shock reflection was observed. The ion $(x, p_x)$ phase space captured at $\omega_0t = 19000$ is plotted in Fig. \ref{fig:Te_a0_over_2}. Hence, the mechanism triggering shock reflection must be linked with the initial density, velocity or electron temperature gradients (Fig. \ref{fig:Te_S3}) imprinted by the laser itself.

\section{Conclusions}
\label{Conclusions}

A parametric 1-D PIC simulations study has been conducted to derive an optimal set of experimental parameters regarding the interaction of a near-critical non-uniform gas profile with a $I_L = 1.7 \times 10^{20}$W/cm$^2$ ($a_0 = 8.8$), $\lambda_L = 0.8 \ \mu$m, $\tau_L = 30$ fs laser pulse. We simulated realizable experimental conditions in order to understand the different ion acceleration mechanisms present during the interaction. At the same time, we identified a range for electron densities in which collisionless electrostatic shock (CES) formation occurs. 

A key factor towards triggering a CES is to have the laser shinning through the profile's density peak while being strongly absorbed (>90$\%$) in the gas up-ramp, where a hot electron population is created. This was achieved with a fourteen times reduction of the original density profile (Fig. \ref{fig:Ey_charts_areal_density}). Once the correct density is achieved, the fast-ion structure originating from TNSA triggered in the density peak evolves into a CES during its propagation across the plasma ramps. This shock can, given the correct conditions, reflect background ions from the down-ramp. CES formation was achieved in the moderate electron density $n_e$ range between 0.35 $n_c$ and 0.7 $n_c$ (for the profile density peak values). IAW formation and background ion reflection, preferentially in the direction of the IAW propagation, was seen at higher densities. At low densities, the ion velocity profile formed by TNSA is too shallow to overtake the ions uniformly accelerated in the quasi-exponential target wings. Shock formation is highly sensible to the target density and to the density gradients inherent to the gas profile. Hence, a controlled and repetitive gas profile production is key towards controlling the laser gas interaction.

The direct contribution of the laser to ion acceleration is that of an initial fast and strong perturbation over the ions, which modulates both the ion and electron phase spaces. Such modulations create and enhance density perturbations across the profile giving rise to strong charge separation fields inside the plasma. However, the electrostatic shock seed lays elsewhere, with the strong electron pressure gradients, which, linked with the profile's density variations, create plasma slabs of different electronic temperatures and densities. The interaction between these plasma slabs leads to the formation of non-linear trapping structures. The latter can transition into electrostatically-reflecting shocks while traversing the density up and down-ramps. Fully ionized ionic species will experience a higher electrostatic potential in the shock region an are therefore more likely to be reflected. The initial "kick" of the laser acts as a multiplier of charge separation and therefore of the electrostatic barriers, giving rise to stronger shock waves than in the laser-free simulation, which will inherently travel faster and hence reflect particles to higher velocities, see ion $(x,p_x)$ phase spaces of Figs. \ref{fig:Final_ion_a0_0}a and b corresponding to the laser-free and laser-on simulations, respectively.

Since the electron pressure gradients are essential for shock formation, the same laser interacting with a different density profile with smoother density gradients would not be as effective as it is in this case. This advantage clearly highlights the interest of employing shock nozzles, which are precisely designed to produce density profiles with strong density gradients.

Given the complexity and number of the laser-matter interaction processes the possibility of performing statistical measures in parametric schemes are fundamental for conducting thorough experimental studies. It is then of vital importance to work at HRR facilities, where $\approx$100 shots can be available each day, in order to acquire good statistics. Such HRR facilities must combine the high laser repetition rate with data flow and data acquisition software and hardware, as well as adapted laser diagnostics and targetry.

\section{Acknowledgments}

This work has received funding from the European Union's Horizon 2020 research and innovation program under grant agreement no 871124 Laserlab-Europe. 

We received financial support from the French State, managed by the French National Research Agency (ANR) in the frame of the Investments for the future Programme IdEx Bordeaux - LAPHIA (ANR-10-IDEX-03-02).

We acknowledge Dr. J.L. Henares for providing the S900 shock nozzle design realized by performing computer fluid dynamic simulations using the Fluent code \cite{Ansys_2018}. We acknowledge Dr. Thanh-ha Nguyen-bui for the complementary CFD simulations performed with the ARES code \cite{Nguyen_2006} inputting the S900 nozzle design.

We acknowledge the support from the LIGHT S\&T Graduate Program (PIA3 Investment fot the Future Program, ANR-17-EURE-0027).

We acknowledge GENCI for providing us access to the supercomputer Irene under the grants no. A0070506129 and no. A0080507594.  

\section*{Data Availability Statement}

The data that support the findings of this study are available from the corresponding author upon reasonable request.

\bibliography{Parametric_study}

\end{document}